\documentclass[twocolumn,preprintnumbers,showpacs,pre,floatfix,amsmath,amssymb,notitlepage,superscriptaddress]{revtex4-1}

\usepackage{bm,bbm}
\usepackage{graphicx,epsfig,subfigure}
\usepackage{amsmath,amssymb}
\usepackage{color,xcolor}
\usepackage{url}
\usepackage[utf8]{inputenc}
\usepackage{enumitem}
\usepackage{setspace}
\usepackage{dcolumn}
\usepackage{bm}
\usepackage{times}
\usepackage{enumerate}
\usepackage{footnote}
\input epsf

\usepackage{bm}
\usepackage{bbm}
\usepackage{graphicx}

\usepackage{xcolor}
\usepackage{enumitem}

\begin{document}

\newtheorem{thm}{Theorem}[section] 
\newtheorem{cor}{Corollary}[section] 
\newtheorem{definition}{Definition}[section]
\newtheorem{remark}[thm]{Remark}
\newtheorem{proof}{Proof}[section]

\newcommand{\new}[1]{{\color{black}#1}}


\title{\new{Multiplex Markov Chains:  Convection Cycles and  Optimality} }

\author{Dane Taylor}
\email{danet@buffalo.edu}
\affiliation{Department of Mathematics, University at Buffalo, State University of New York, Buffalo, NY 14260, USA}

\date{\today} 

\begin{abstract}
Multiplex networks are a common modeling framework for interconnected  systems and multimodal data, yet we still lack fundamental insights for how multiplexity affects  stochastic processes.  We introduce  a \new{novel} ``Markov chains of Markov chains'' model \new{called \emph{multiplex Markov chains} (MMCs)} such that with  probably $(1-\omega)\in[0,1]$ random walkers remain in the same layer and follow (layer-specific) \emph{intralayer Markov chains}, whereas  with probability $\omega$ they move to different layers following (node-specific) \emph{interlayer Markov chains}.  \new{One} main finding is the identification of
\emph{multiplex convection}, whereby a stationary distribution exhibits circulating flows that involve multiple layers. Convection cycles are well understood in fluids, but   \new{are insufficiently explored on networks.} 
\new{Our experiments reveal that one mechanism for  convection is the existence of imbalances for the (intralayer) degrees of   nodes in different layers. To gain further  insight, we employ spectral perturbation theory to characterize the stationary distribution  for the limits of small and large $\omega$, and we show that MMCs inherently exhibit optimality for intermediate $\omega$ in terms of their convergence rate and the extent of convection. As an application, we conduct an MMC-based analysis of    brain-activity data, finding    MMCs  to differ between healthy persons and those with Alzheimer's disease. 
Overall, our work suggests MMCs and convection as two important new directions for network-related  research.}
\end{abstract}

\maketitle

\new{\section{Introduction}}

\emph{Convection cycles}  are a well-understood phenomenon in fluid dynamics \cite{falkovich2011fluid}, but they remain underexplored in the context of networks.
While convection traditionally arises   under external forces such as buoyancy,  here \new{we find it to}  emerge  as a multiplexity-induced phenomenon \new{for multiplex Markov chains (MMCs), which is a modeling framework  that can be applied to study  diffusion on multiplex networks}. 
%
Multiplex networks \cite{cozzo2018multiplex}---a type of multilayer network  \cite{boccaletti2014,kivela2014} in which layers  encode different types of \emph{intralayer} edges, and \emph{interlayer} edges couple  the layers---have  been used to model interconnection complex systems including transportation networks \cite{taylor2015topological,strano2015multiplex,sole2019effect},  critical infrastructures \cite{haimes2001leontief}, and different types of   relationships  \cite{krackhardt1987cognitive}. 
They \new{also} provide frameworks for data integration/fusion \cite{de2015structural,taylor2016enhanced,taylor2016_b} and stratification \cite{guillon2017loss,soriano2018spreading}.

Here,  we propose a multiplex generalization of Markov chains  \cite{kemeny1976markov}, a memoryless process for stochastic transitions between discrete states that provides   a theoretical foundation for diverse applications, such as   queuing theory \cite{kendall1953stochastic},  population dynamics \cite{kingman1969markov}, and machine-learning algorithms that rely on  Markov chain Monte Carlo \cite{gilks1995markov}, hidden Markov models \cite{tierney1994markov}, and/or  Markov decision processes \cite{parr1998reinforcement}. 
\new{(See also Markov stability \cite{delvenne2010stability,schaub2012markov} for multiscale community detection.)}
Similar to Markov chains, MMCs \new{will} find diverse applications within, \emph{and beyond},  the study of networks.

According to an MMC,  random walkers move along (layer-specific) \emph{intralayer Markov chains} with probability $\omega\in[0,1]$, and with probability $1-\omega$ they  transition to new layers following (node-specific) \emph{interlayer Markov chains}.  MMCs  can be used to study random walks on multiplex networks,  and diffusion physics for multiplex networks is already a burgeoning field \cite{mucha2010,gomez2013diffusion,sole2013spectral,radicchi2013abrupt,de2014navigability,tejedor2018diffusion,cencetti2019diffusive}.  Most approaches rely on a generalization of the graph Laplacian called a supraLaplacian matrix, which can be constructed by first multiplexing the network layers' adjacency matrices into a supra-adjacency matrix, and then creating a Laplacian matrix by treating the supra-adjacency matrix as if it were a normal adjacency matrix (i.e., neglecting that  intra-- and interlayer edges are different). Both  normalized \cite{mucha2010} and unnormalized supraLaplacians \cite{gomez2013diffusion,sole2013spectral,radicchi2013abrupt,de2014navigability,tejedor2018diffusion,cencetti2019diffusive} have been studied, and in the latter case, one can simply couple the unnormalized Laplacians of layers. Other formulations \new{for diffusion on multiplex networks} have also been proposed to study centrality and consensus \cite{trpevski2014discrete,dedom2015,sole2016random,taylor2017eigenvector,deford2018new,taylor2019supracentrality,taylor2019tunable}. Despite the  significant advances that have been made, this field remains in its infancy \cite{de2016physics}.

\begin{figure*}[t]
\centering
\includegraphics[width=\linewidth]{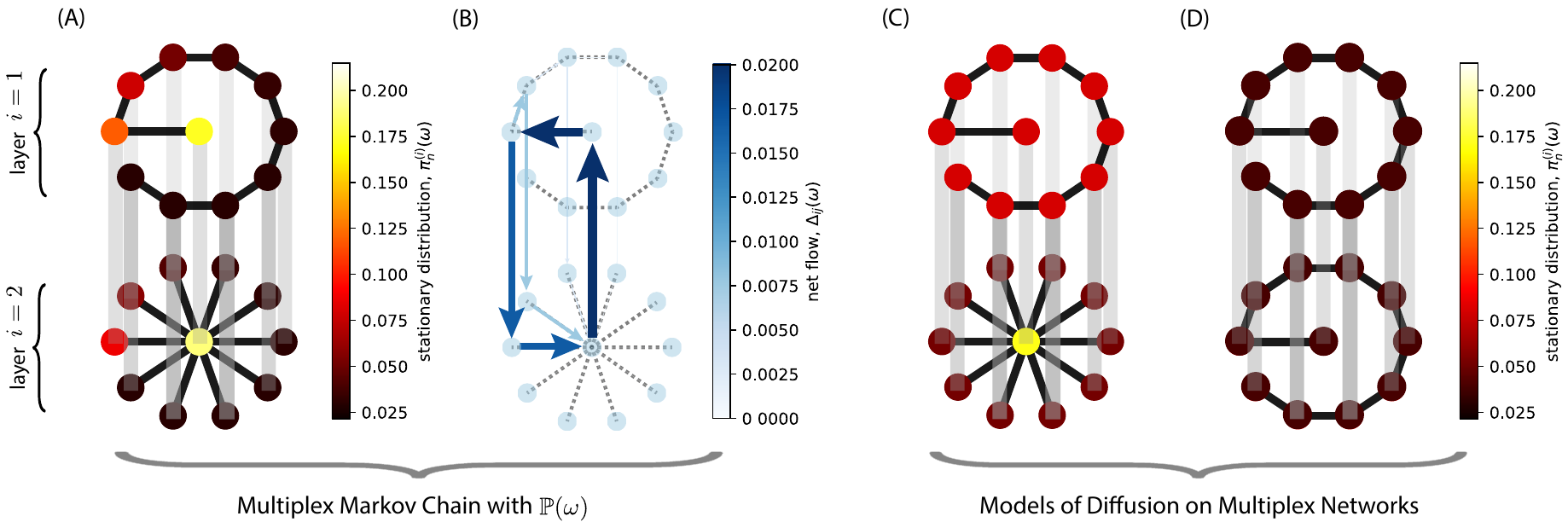}
\caption{{\bf \new{Multiplex Markov chains (MMCs) versus diffusion models for multiplex networks}.}
(A)~Stationary distribution $\pi_{n}^{({i})}(\omega)$ for a discrete-time MMC with  interlayer coupling strength $\omega=0.5$.   The   black and   grey lines indicate edges in the intra-- and interlayer Markov chains, respectively.
(B) \new{The} stationary net flow $\Delta_{pq}(\omega) =  \pi_{n_p}^{({i_p})}(\omega) [\mathbb{P}(\omega)]_{pq} - \pi_{n_q}^{({i_q})}(\omega) [\mathbb{P}(\omega)]_{qp}$ \new{reveals for each edge $(p,q)$ the extent to which diffusion moves in a biased direction after the system converges to its steady state}.  Observe the emergent  \emph{multiplex convection cycle}, which  \new{is a cyclical flow at stationarity that involves more than one layer. It} arises due to \emph{multiplex imbalance}, whereby the multiplexing of  reversible Markov chains   yields an irreversible MMC.
\new{See Sec.~\ref{sec:convection} for more detail.
(C)--(D) The  stationary distributions of models for diffusion on multiplex networks that first couple layers into a supra-adjacency matrix $\hat{\mathbb{A}}(\omega)$  do not exhibit convection cycles nor do they reflect global properties of the multiplex network (see Appendix A). (C) The stationary distribution of a supratransition matrix $\hat{\mathbb{P}}(\omega) = \hat{\mathbb{D}}(\omega)^{-1}\hat{\mathbb{A}}(\omega)$ \cite{mucha2010,sole2016random} is determined by the \emph{total} degrees, which are encoded in the diagonal matrix $\hat{\mathbb{D}}(\omega)$.
(D) The stationary distribution of an unnormalized supraLaplacian $\hat{\mathbb{L}}(\omega) = \hat{\mathbb{D}}(\omega) - \hat{\mathbb{A}}(\omega)$ \cite{gomez2013diffusion,sole2013spectral,radicchi2013abrupt,de2014navigability,tejedor2018diffusion,cencetti2019diffusive} is the uniform distribution. 
}
}
\label{fig:toy1}
\end{figure*}

By coupling  ``Markov-chain layers'' rather than ``network layers,'' we  identify and study {a} novel multiplexity-induced phenomenon called 
 \emph{multiplex convection},  which---along with the convergence rate $\lambda_2$---is  found to be optimized  at intermediate $\omega$.  These properties are shown to have an interesting and complicated relation to the imbalances of nodes' intralayer degrees. We analyze MMCs  with spectral perturbation theory to characterize the stationary distribution when there is a separation of timescales between intra-- and interlayer transitions: As $\omega\to0$,  intralayer Markov chains each  approach (local) stationary solutions, and these (layer-specific) solutions are balanced by the interlayer Markov chains.  Analogously, as $\omega\to1$  the interlayer Markov chains individually  approach (local) stationary solutions, and these (node-specific) solutions are balanced by the intralayer Markov chains.  
\new{As an application, we study frequency-multiplexed brain data through the lens of MMCs, highlighting differences between the brains of healthy persons and those with Alzheimer's disease.}  

\new{This paper is organized as follows: 
We define MMCs in Sec.~\ref{sec:methods} and present our theory in  Sec.~\ref{sec:theory}.
In Sec.~\ref{sec:optimal}, we study the  optimality of MMCs. 
In Sec.~\ref{sec:brain}, we   analyze  brain-activity data. 
We present a summary in Sec.~\ref{sec:conclusion}. Code for the experiments is available at \cite{daneGithub}.}

 
\new{\section{Multiplex Markov chains}\label{sec:methods}}

\new{\subsection{Model}\label{sec:model}}

Consider a set of  \emph{intralayer Markov chains} (the ``layers'') with size-$N$ transition matrices $ {\bf P}^{(i)} $ for $i\in\{1,\dots,I\}$ and a set of (node-specific) \emph{interlayer Markov chains} with size-$I$  transition matrices $\tilde{\bm{P}}^{(n)}$ for $n\in\{1,\dots,N\}$. Each $ {P}^{(i)}_{nm}$  scales the transition probability from node $n$ to $m$ in layer $i$, while $ \tilde{{P}}_{ij}^{(n)}$ scales the transition probability from layer $i$ to $j$ for node $n$. Or equivalently, these give transitions  between node-layer pairs:  from $(n,i)$ to $(m,i)$ in the case of $ {P}^{(i)}_{nm}$, and from $(n,i)$ to $(n,j)$ in the case of $ \tilde{P}^{(n)}_{ij}$.


A (discrete-time) multiplex Markov chain (MMC) is a stochastic process in which the states are node-layer pairs $(n_p,i_p)$, which we enumerate $p\in\{1,\dots, NI\}$, yielding  $n_p= (p \textrm{ mod } N)$ and   $i_p=\lceil p/N \rceil$ ( `mod' and $\lceil \cdot \rceil$ indicate the modulus and ceiling functions, respectively.) Transitions between states follow a \emph{supratransition matrix} 
 \begin{align}
\mathbb{P}(\omega) &= (1-\omega) \textrm{diag}\left(\{{\bf P}^{(i)}\}\right)
+  \omega \sum_n  \tilde{\bm{P}}^{(n)} \otimes  {\bf E}^{(n)}.
\label{eq:supra_t}
\end{align}
\emph{Coupling strength} $\omega\in[0,1]$ tunes  the probability that random walks use interlayer vs intralayer Markov chains, $\textrm{diag}(\cdot)$ is a block-diagonal matrix in which the argument matrices are placed along the diagonal, $\otimes$ denotes the Kronecker product, and ${\bf E}^{(n)}=  {\bf e}^{(n)}[{\bf e}^{(n)}]^T$, where   ${\bf e}^{(n)}_m = 1$ if $n=m$ and 0 otherwise. Each   $[\mathbb{P}(\omega)]_{pq}$ gives the transition probability from $(n_p,i_p)$ to $(n_q,i_q)$. Under the assumption of \emph{uniform coupling}---\new{that is, all interlayer Markov chains are identical---, }  we let $\tilde{\bm{P}}^{(n)}=\tilde{\bm{P}}~\forall n$, and the last term in Eq.~\eqref{eq:supra_t} simplifies to $\omega \tilde{\bm{P}} \otimes {\bf I}$, where ${\bf I}$ is a size-$N$ identity matrix.


Let $\mathbbm{x}^{(t)}$  be a length-$NI$ row vector such that $[\mathbbm{x}^{(t)}]_{p}$  gives the expected fraction of random walkers at node-layer pair $(n_p,i_p)$ at time $t$. Given initial condition $\mathbbm{x}^{(0)}$,   $\mathbbm{x}^{(t)}$ evolves following a linear discrete map 
\begin{align}\label{eq:ahhs}
\mathbbm{x}^{(t+1)} = \mathbbm{x}^{(t)} \mathbb{P}(\omega).
\end{align} 
If $\mathbb{P}(\omega)$ is nonnegetive, irreducible, and aperiodic, then $\mathbbm{x}^{(t)}\to \mathbbm{v}(\omega)$  converges to a  stationary distribution, 
which is the left dominant eigenvector of  $\mathbb{P}(\omega)$. It is convenient to define $\pi_{n_{p}}^{({i_{p}})}(\omega) = [\mathbbm{v}(\omega)]_p$ and to drop the subscript $p$, so that  $\pi_{n}^{({i})}(\omega)$ is the density of walkers at node $n$ in layer $i$.

We also define a (continuous-time) MMC with a normalized supraLaplacian 
\vspace{-.2cm}
\begin{align}
\mathbb{L}(\omega) = \mathbb{I} - \mathbb{P}(\omega) ,
\end{align}
\new{where} $\mathbb{I}$ is a size-$(NI)$ identity matrix. \new{Entries}   $[\mathbb{L}(\omega) ]_{pq}$  are   rates for  transitions between node-layer pairs. \new{Letting $\mathbbm{x}(t)$ denote the distribution of random walkers at time $t$, it   evolves in time as 
$\frac{d}{dt}\mathbbm{x}(t) = -\mathbb{L}(\omega )^T\mathbbm{x}(t)$.  The stationary distribution for this process is identical to that of Eq.~\eqref{eq:ahhs}. One can  also define a consensus/synchronization process  \cite{skardal2014optimal,taylor2016synchronization} by $\frac{d}{dt}\mathbbm{x}(t) = -\mathbb{L}(\omega )\mathbbm{x}(t)$, for which the stationary distribution   is the  uniform distribution.
 For the remainder of this paper, we will focus   on}   discrete-time MMCs due to the wealth of existing knowledge on supraLaplacians \cite{mucha2010,gomez2013diffusion,sole2013spectral,radicchi2013abrupt,de2014navigability,tejedor2018diffusion,cencetti2019diffusive}.


\new{\subsection{Application of MMCs to  multiplex networks}\label{sec:app} }

\new{Although a MMC need not be constructed from a multiplex network,}
MMCs \new{do} provide a new form of diffusion on multiplex networks. 
Let ${\bf P}^{(i)}=[{\bf D}^{(i)}]^{-1}{\bf A}^{(i)}$ and  $\tilde{\bm{ P}}^{(n)} =[\tilde{\bm{ D}}^{(n)}]^{-1}\tilde{\bm{ A}}^{(n)}$ denote  intra-- and interlayer transition matrices, respectively, of Markov chains derived from a multiplex network in which ${\bf A}^{(i)}$ and $\tilde{\bm{A}}^{(n)}$ are   intra-- and interlayer adjacency matrices, and ${\bf D}^{(i)}$ and $\tilde{\bm{ D}}^{(n)}$ are diagonal matrices with entries  $\new{{D}^{(i)}_{nn} =  d_n^{(i)}} = \sum_m A_{nm}^{(i)} $ and $\new{\tilde{{D}}_{ii} = \tilde{d}_i^{(n)}= }\sum_j \tilde{A}^{(n)}_{ij}$ that encode the \new{\emph{intralayer degrees} and \emph{interlayer degrees}.}

\new{Before continuing, we note that there may be applications in which the transition matrices  are not constructed from adjacency matrices. That is, in general an inter-- or intralayer transition matrix does not necessary have take the specific functional form $P = D^{-1}A$. \emph{We therefore highlight that the study of diffusion on multiplex networks is just one of  many potential applications for MMCs.} }

In Fig.~\ref{fig:toy1}(A), we visualize  an MMC in which ${\bf P}^{(1)}$ and ${\bf P}^{(2)}$ are intralayer transition matrices associated with chain and star networks (both are undirected and unweighted). Self-edges are added to the first and last nodes of the chain to make it 2-regular. We enumerate the nodes $n\in\{1,\dots,11\}$ clockwise around the chain, starting at the center node. We implement uniform coupling with $\tilde{\bm{P}}  = \left[{{0}\atop{1  }} {{1  }\atop{ 0}} \right]$.
Node colors in Fig.~\ref{fig:toy1}(A) depict   $\pi_n^{(i)}(\omega)$ for  $\omega=0.5$. 

Note that $\pi_n^{(i)}(\omega)$ is largest for node-layer pair $(n,i) =(1,2)$ (the hub node in the star network); $\pi_n^{(i)}(\omega)$ is also large for node-layer pair $(1,1)$, since it is coupled to $(1,2)$ by an interlayer edge. \new{Also,}
observe for layer $i=1$ that the $\pi_n^{(i)}(\omega)$ values monotonically decrease as one moves clockwise around   the chain.
That is, the \new{stationary distributions} of MMCs are influenced by    global structure, which can be beneficial, for example, if one seeks to study the importances of nodes and/or layers  
\cite{trpevski2014discrete,dedom2015,sole2016random,taylor2017eigenvector,deford2018new,taylor2019supracentrality,taylor2019tunable}. 

%

{As shown in Figs.~\ref{fig:toy1}(C)--(D), MMCs provide an important contrast to popular models for diffusion on multiplex networks in which one first couples the layers' adjacency matrices into a supra-adjacency $\hat{\mathbb{A}}(\omega)$, and then one subsequently defines a diffusion process that treats $\hat{\mathbb{A}}(\omega)$ as if it were a standard adjacency matrix of a single-layer network. Note that this step neglects that  inter-- and intralayer edges are different types of edges. We describe these models in detail in  Appendix A and provide a brief summary here.  This approach can give rise to a different type of supratransition matrix    \cite{sole2016random} $\hat{\mathbb{P}}(\omega) = \hat{\mathbb{D}}(\omega)^{-1}\hat{\mathbb{A}}(\omega)$, a different normalized supraLaplacian matrix  
$  {\mathbb{I}}  - \hat{\mathbb{P}}(\omega)$ \cite{mucha2010}, and an unnormalized supraLaplacian $\hat{\mathbb{L}}(\omega) = \hat{\mathbb{D}}(\omega) - \hat{\mathbb{A}}(\omega)$ \cite{gomez2013diffusion,sole2013spectral,radicchi2013abrupt,de2014navigability,tejedor2018diffusion,cencetti2019diffusive}. (Here, $\hat{\mathbb{D}}(\omega) $ is a degree matrix and ${\mathbb{I}} $ is the identity matrix.)
We find that these models do not give rise to emergent convection cycles, due in part to the fact that their stationary distributions do not reflect global properties of a multiplex network:
For $\hat{\mathbb{P}}(\omega)$ and $ {\mathbb{I}}-\hat{\mathbb{P}}(\omega)$,  the stationary distribution is proportional to the degrees. For $\hat{\mathbb{L}}(\omega)$, it is the uniform distribution. 
}


\pagebreak

\new{\subsection{Convection cycles in MMCs}\label{sec:convection}} 

The $\pi_n^{(i)}(\omega)$ values reflect a complicated interplay between intra-- and interlayer Markov chains, which we further study through  the stationary flows \new{across edges}
\begin{align}
\mathbb{F}_{pq}(\omega) = \pi_{n_p}^{({i_p})}(\omega)[\mathbb{P}(\omega)]_{pq}.
\end{align}
We further define $\mathbb{F}(\omega) = \overline{\mathbb{F}}(\omega) + \Delta(\omega)/2$ to
separate the matrix  into its symmetric part, $[\mathbb{F}(\omega) + \mathbb{F}(\omega)^T]/{2}$, and skew-symmetric part,  $ {\Delta }(\omega)/{2}$. Each entry
\begin{align}
\Delta_{pq}(\omega)  =  \mathbb{F}_{pq}(\omega)   - \mathbb{F}_{qp}(\omega)
\end{align}
indicates the  \emph{stationary  flow imbalance}  across each edge.  $\Delta_{pq}(\omega)>0$  implies that there is a greater flow from $(n_p,i_p)$ to $(n_q,i_q)$, whereas $\Delta_{pq}(\omega)<0$ implies the opposite. To shorten our notation, in the rest of the paper we will drop the argument $\omega$ in $\mathbb{F}(\omega)$ and $\Delta(\omega)$.

 We define the \emph{total flow imbalance}  $||\Delta||_F$ using the Frobenius norm, and a Markov chain is  \emph{reversible} iff  $||\Delta||_F=0$ (i.e., $\Delta_{pq}=0~\forall p,q$). We define \emph{multiplex imbalance} as the phenomenon whereby the multiplex coupling of reversible Markov chains yields an irreversible MMC. In that case,  $||\Delta||_F$ quantifies the \emph{total multiplex imbalance}.

In Fig.~\ref{fig:toy1}(B), we visualize $\Delta_{pq}$  for the MMC from Fig.~\ref{fig:toy1}(A). Observe that these values  exhibit a circulating flow imbalance involving more than one layer, which we call \emph{multiplex convection}.  Because ${\bf P}^{(1)}$,  ${\bf P}^{(2)}$ and $\tilde{\bm{P}}$ are all transition matrices of reversible Markov chains, the irreversibility of the MMC is an emergent multiplexity-induced property. 

\new{Note}
that  $\Delta_{pq}$ is largest from node-layer pair $(1,2)$ to $(1,1)$, and this edge is also associated with the largest imbalance of intralayer degrees: $d_{1}^{(2)}  =10$, $d_{1}^{(1)}=2$, and $d_{1}^{(2)} -  d_{1}^{(1)} =  8$ (recall the chain layer has self-edges to make it 2-regular).
\new{This reveals an important}
mechanism that contributes to the emergence of multiplex imbalance and convection:   $\Delta_{pq}$ is often large for an interlayer edge associated with a node $n$ such that its intralayer degrees $d_{n}^{({i_p})} $ and $d_{n}^{({i_q})} $ are imbalanced (where $d_{n}^{({i})}  = \sum_m A_{nm}^{(i)}$).
%

~
 
\new{\section{Timescale separation analysis}\label{sec:theory}}

We   analyze  $\pi_{n}^{({i})}(\omega)$  for two limits:
\new{as} $\omega\to 0$: \new{random}  walkers rarely move between layers, yielding a type of layer decoupling.
\new{As} $\omega\to 1$: \new{random} walkers rarely remain in the same layer,  yielding  a type of layer aggregation. 
\new{To this end, we develop spectral perturbation  theory in Appendix B. Here, we will   summarize these mathematical results.}

Given intra-- and interlayer Markov chains with transition matrices ${\bf P}^{(i)} $ and $\tilde{\bm{ P}}^{(n)}$, respectively, let ${\bf v}^{(i)}$ and  $\tilde{\bm{ v}}^{(n)}$ denote  their stationary distributions. Note that they are   left eigenvectors associated with an eigenvalue equal to 1. Let $\tilde{\bm{ e}}^{(i)}$ be a length-$I$ unit vector (i.e., $\tilde{e}_j^{(i)}=1$ if $j=i$ and 0 otherwise).
%
%
%
%
%
%
%
%
%
Setting $\omega=0$ in Eq.~\eqref{eq:supra_t} yields $\mathbb{P}(0) =  \textrm{diag}\left(\{{\bf P}^{(i)}\}\right)$, for which $\lambda_1=1$ is an eigenvalue. \new{With  Theorem~B.1 in Appendix B, we show that $\lambda_1$ has}  an $I$-dimensional left eigenspace spanned by vectors  
\begin{align}
\mathbbm{v}^{(i)} = \tilde{\bm{ e}}^{(i)}\otimes {\bf v}^{(i)}.
\end{align}
However, for any positive $\omega$,
Perron-Frobenius theory 
 \cite{bapat1997} ensures that the eigenvalue $\lambda_1=1$ of $\mathbb{P}(\omega)$ has a 1-dimensional eigenspace spanned by a unique left dominant eigenvector $\mathbbm{v}(\omega)$. Moreover, 
 $\mathbbm{v}(\omega)$ {must}   converge within the  subspace   $\textrm{span}(\mathbbm{v}^{(i)} )$, implying   there exist constants $\tilde{\alpha}_i$ such that $\lim_{\omega\to 0} \mathbbm{v}(\omega) = \sum_i \tilde{\alpha}_i\mathbbm{v}^{(i)}. $ We denote $ \tilde{\bm{\alpha}} = [ \tilde{\alpha}_1,\dots, \tilde{\alpha}_I]$. 

\new{With  Theorem~B.3 in Appendix B,} we show that  $ \tilde{\bm{\alpha}}$ is the dominant left eigenvector of an ``effective'' interlayer transition  matrix $\tilde{\bm{ X}} $ with entries $\tilde{X}_{ij}= \sum_n {v}_n^{(i)} \tilde{  { P}}_{ij}^{(n)}$.  Using the notation $[\mathbbm{v}(\omega)]_p = { {\pi}}_{n_p}^{({i_p})}(\omega) $, we have 
\begin{align}\label{eq:lim0}
\lim_{\omega\to0} { {\pi}}_n^{(i)}(\omega)  &=   \tilde{\alpha}_{i} v_n^{(i)}.
\end{align}
That is, each intralayer Markov chain $i$ obtains its   stationary distribution  ${\bf v}^{(i)}$, and these local  solutions are balanced by the stationary distribution of an ``effective'' interlayer Markov chain that depends  on all inter-- and intralayer Markov chains. In the case of uniform coupling, $\tilde{\bm{P}}^{(n)}=\tilde{\bm{P}}$,  $\tilde{\bm{ X}} =\tilde{\bm{P}}$, and   $ \tilde{\bm{\alpha}} =\tilde{\bm{v}} $ (the left dominant eigenvector of $\tilde{\bm{P}}$).

We  analyze the limit $\omega\to1$ in a similar way, except we first implement a change of basis via the (unitary) stride-permutation matrix $\mathbb{U}$ that reorders the node-layer pairs as layer-node pairs: 
$ [\mathbb{U}]_{pq} = 1$ if $q=\lceil p/N \rceil+T\,[(p-1)\bmod N]$ and  $[\mathbb{U}]_{pq} = 0$ otherwise. Matrix $\mathbb{Q}(\omega) = \mathbb{U}\mathbb{P}(\omega)\mathbb{U}^T$ is a supratransition matrix for the same MMC as $\mathbb{P}(\omega)$; the only difference is that the rows and columns have been permuted. We obtain
\begin{align}
\mathbb{Q}(\omega) = (1-\omega) \sum_i  {{\bf P}}^{(i)} \otimes  \tilde{\bm{ E}}^{(i)}  + \omega ~\textrm{diag}\left(\{\tilde{\bm{ P}}^{(n)}\}\right),
\end{align}
where $\tilde{\bm{ E}}^{(i)} = \tilde{\bm{e}}^{(i)}[\tilde{\bm{e}}^{(i)}]^T$ and $\tilde{\bm{e}}^{(i)}$ is a length-$I$ unit vector. Observe that the form of  $\mathbb{Q}(\omega)$  qualitatively matches that of Eq.~\eqref{eq:supra_t}; only the inter-- and intralayer transition matrices have been swapped. Thus,  one can equally interpret an MMC as intralayer Markov chains coupled by intralayer ones, or as interlayer Markov chains coupled by intralayer ones. These are formally the same. We can thus make use of our earlier results to obtain
\begin{align}
\lim_{\omega\to1} { {\pi}}_n^{(i)}(\omega)  &= {\alpha}_{n}   \tilde{v}_{i}^{(n)}   ,
\label{eq:lim1}
\end{align}
where $\bm{ \alpha}=[\alpha_1,\dots,\alpha_N] $ is the dominant left eigenvector  of a transition matrix ${\bf X}$ with entries $ X_{nm} = \sum_i   \tilde{v}_{i}^{(n)} {P}_{nm}^{(i)} $ for an ``effective'' intralayer Markov chain.

\begin{figure}[t]
\centering
\includegraphics[width=\linewidth]{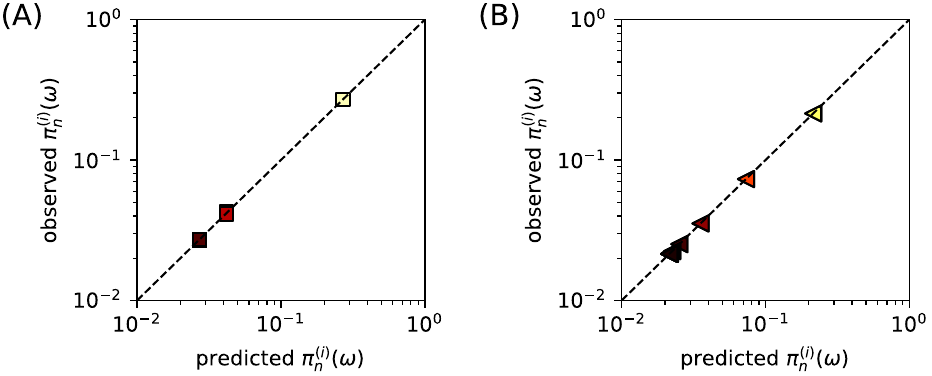}
\caption{
{\bf Validation of theory for timescale separations.}
We compare observed and predicted values of $\pi_n^{(i)}(\omega)$ 
for the MMC from Fig.~\ref{fig:toy1} for   two limiting cases:
(A)~Eq.~\eqref{eq:lim0} for $\omega\to 0$; and 
(B)~Eq.~\eqref{eq:lim1} for $\omega\to 1$.
The symbols' colors correspond to the same color scale as that shown in  Fig.~\ref{fig:toy1}(A). Note that 22 symbols are plotted in each panel, but because they overlap, we perceive just a few.
}
\label{fig:asym}
\end{figure}

In Fig.~\ref{fig:asym}, we validate the accuracy of Eqs.~\eqref{eq:lim0} and \eqref{eq:lim1} for the MMC shown in Fig.~\ref{fig:toy1}. The observed $\pi_n^{(i)}(\omega) $ values for small and large $\omega$    were computed with $\omega=10^{-3}$ and  $\omega = 1-10^{-3}$.

\new{
Equations~\eqref{eq:lim0} and \eqref{eq:lim1} have the important consequence of implying} that  $\lambda_2(\omega)$---the second-largest-in-magnitude eigenvalue  of $\mathbb{P}(\omega)$---is  optimized  at   {some intermediate}  value of $\omega$. Because $\mathbbm{x}^{(t)}_p\to {\pi}_{n_p}^{(i_p)}(\omega)$ with  $t\to\infty$ as $\mathcal{O}(\lambda_2^t(\omega))$, $\lambda_2(\omega)\in(0, 1)$ is called the  \emph{convergence rate}. Importantly, because 
$\mathbb{P}(w)$   {has} $I$-dimensional and $N$-dimensional dominant eigenspaces when $\omega=0$ and $\omega=1$, respectively, 
it follows that  $\lambda_2(\omega)\to1$ in either limit.  {Finally},  Rolle's theorem \cite{sahoo1998mean}  {implies there is}  a  minimum   {since} $d \lambda_2(\omega) / d\omega<1$ at $\omega=0$ and $d \lambda_2(\omega) / d\omega>1$ at $\omega=1$.
 \new{We  study this optimality for MMCs, as well as another type of optimality, in the next section.}

~

\new{\section{Optimality of MMCs for intermediate $\omega$}\label{sec:optimal}} 

Next, we show that  total multiplex imbalance $||\Delta||_F$ is maximized at some value $\omega^*_{ \Delta} = \text{argmax}_\omega ||\Delta||_F\in(0,1)$. We use the same interlayer Markov chains as in Fig.~\ref{fig:toy1}, but now allow the  interlayer Markov chains to be different for every node:
 $\tilde{\bm{P}}^{(n)}  = \left[{{(1-a_n) }\atop{a_n  }} {{a_n  }\atop{ (1-a_n)}} \right]$, where $a_n\in[a,1]$ tunes the probability of switching layers at each node $n$.  
We consider four strategies for choosing  $a_n$:
\vspace{.2cm}
\new{\begin{itemize}[nosep]
\item[(I)] Identical $a_n$: we define $a_n=a$ for each $n$;
\item[(II)] Increasing $a_n$: we define $a_n = a + (n-1)\delta a$ for $n\in\{1,\dots,N\}$ with $\delta a= (1-a)/(N-1)$; 
\item[(III)] Decreasing $a_n$: we let $a_n = 1 - (n-1)\delta a$;  
\item[(IV)] Random $a_n$: we sample $a_n$ uniformly at random from $[a,1]$.
\end{itemize}
\vspace{.2cm}
}

\begin{figure*}[t]
\centering
\includegraphics[width=.6\linewidth]{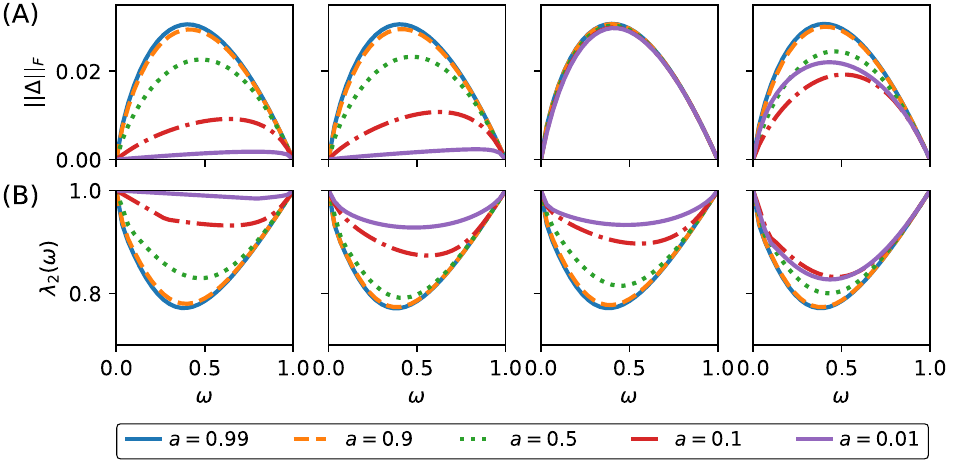}
\caption{
{\bf Optimal imbalance and  convergence rate.}
(A)   Total multiplex imbalance $||\Delta||_F$ and (B) convergence rate $\lambda_2(\omega)$  versus $\omega$ for the same intralayer Markov chains as in Fig.~\ref{fig:toy1}, but with node-specific interlayer Markov chains in which $a_n$ tunes the probability of switching layers at node $n$.
We choose $a_n\in[a,1]$ via four strategies:
(left)~Identical $a_n$: $a_n=a~\forall~n$;
(center-left) Increasing $a_n$: evenly spaced and monotonically increasing with $n$; and 
(center-right) Decreasing $a_n$: evenly spaced and monotonically decreasing with $n$;
(right) Random $a_n$: sampled uniformly at random. We define $\omega^*_{\Delta} = \text{argmax}_\omega ||\Delta||_F$ and $\omega^*_{\lambda_2} = \text{argmax}_\omega \lambda_2(\omega)$. 
}
\label{fig:opt1}
\end{figure*}

\begin{figure*}[t]
\centering
\includegraphics[width=.6\linewidth]{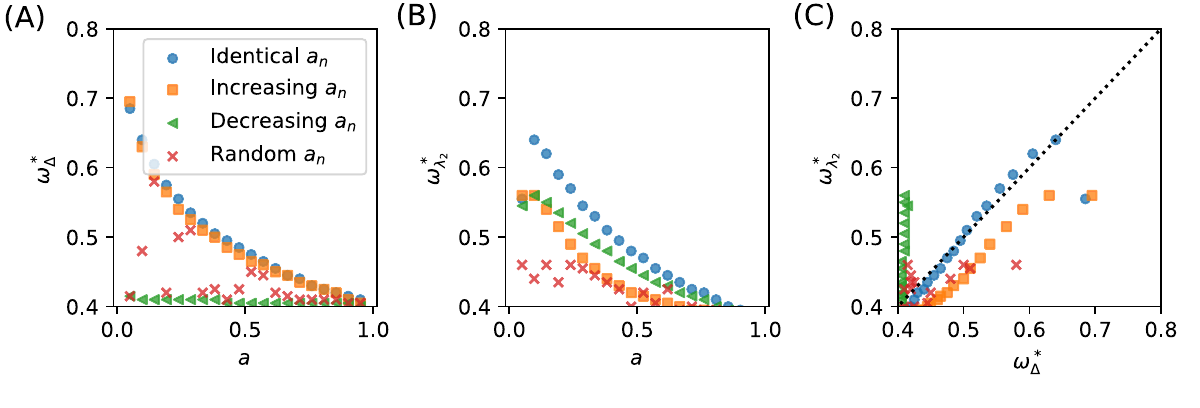}
\vspace{-.6cm}
\caption{
{\bf Relationship between optimum for multiplex imbalance and \new{the optimum} for  convergence rate.}
We show results for  hetergeneous interlayer Markov chains in which the probably to move to a different layer at each node $n$ is tune\new{d} by $a_n$, and we study four strategies for selecting $a_n$, each of which depends on a   $a$.
Panels (A) and (B) depict $\omega_\Delta^*= \text{argmax}_\omega ||\Delta||_F$ and  $\omega_{\lambda_2}^*= \text{argmax}_\omega \lambda_2(\omega)$, respectively, as a function of parameter $a$.
(C) A direct comparison of $\omega_\Delta^*$ and $\omega_{\lambda_2}^*$.
}
\label{fig:opt2}
\end{figure*}

In Fig.~\ref{fig:opt1}(A), we plot $||\Delta||_F$ versus $\omega$; each panel depicts a different strategy. 
First, because $a\to1$ recovers the interlayer transition matrix used in Fig.~\ref{fig:toy1} for all strategies, the (blue solid) curves for $a=0.99$  are nearly identical in all top panels. As $a$ decreases, the different strategies yield different $||\Delta||_F$ curves: (I)--(II) the curves for identical and increasing $a_n$ flatten as $a$ decreases and the location of the optimum shifts toward larger $\omega$; (III) the $||\Delta||_F$ curves for   decreasing $a_n$   are insensitive to $a$; and (IV) the curves for random $a_n$ seem to change randomly, but generally decrease.
These responses can be understood by noting that $a_1$ strongly varies with $a$ for the first two strategies, it remains unchanged for the decreasing-$a_n$ strategy, and it is random for the last strategy, although its expectation   decreases. Parameter $a_1$ determines the optimality of $||\Delta||_F$ in this case, because the net flow $\Delta_{pq}$ is largest from node-layer pair $(1,2)$ to $(1,1)$ [see Fig.~\ref{fig:toy1}(B)], and $a_1$ tunes the probability of walkers make this transition.


%
%
%

In Fig.~\ref{fig:opt1}(B), we show $\lambda_2(\omega)$ versus $\omega$ for the same MMC as in Fig.~\ref{fig:opt1}(A).
Interestingly, the locations $\omega^*_{\Delta}$ and $\omega^*_{\lambda_2}$ of optima for $||\Delta||_F$ and $\lambda_2(\omega)$ appear to be strongly correlated for some strategies. \new{We now explore this further by repeating this experiment with many values of $a\in(0,1)$.}

In  Fig.~\ref{fig:opt2}(A) and (B), we plot $\omega_\Delta^*$ and $\omega_{\lambda_2}^*$, respectively, as a function of $a$. In each panel, we show results for the four strategies for creating interlayer Markov chains. 
Observe that in the limit $a\to1$, all of the optimums occur at approximately the same coupling strength $\omega_{\lambda_2}^*\approx \omega_\Delta^*\approx 0.4$, which is expected since all four strategies yield uniform coupling: $\bm{\tilde{P}}^{(n)}\to\bm{\tilde{P}}   = \left[{{0}\atop{1  }} {{1  }\atop{ 0}} \right].$ 
As one decreases $a$,  the optimums shift to  larger values of $\omega$ for all strategies except for  the strategy with decreasing $a_n$.  This response is largest for the strategies with identical $a_n$ and increasing $a_n$, and it is less clear  for the strategy with random $a_n$ (in which case, the dependence of $\omega_\Delta^*$ on $a$ is more random).

In   Fig.~\ref{fig:opt2}(C), we  compare  $\omega_{\lambda_2}^*$ and $\omega_\Delta^*$ across these values of $a$.
Note for the strategy of identical $a_n$, that the two the optimums occur at nearly the same value of $\omega$ for any $a\in(0,1)$---that is, the blue squares lie along the diagonal. The  dependence of $\omega_{\lambda_2}^*$ and $\omega_\Delta^*$ on $a$ is also strongly correlated for the strategy of increasing $a_n$; however $\omega_\Delta^*$ is always slightly larger than $\omega_{\lambda_2}^*$ and the relationship appears to be nonlinear for small $a$. The relation between  $\omega_{\lambda_2}^*$ and $\omega_\Delta^*$ appears to be random for the random strategy. Interestingly, for the strategy of decreasing $a_n$,  $\omega_{\lambda_2}^*$ clearly increases with decreasing $a$, however  $\omega_\Delta^*$ appears to not depend on \new{$a$}.

\new{We give the following interpretation to provide intuitive insight into these results.}
Recall from Fig.~1(B)  that the imbalance $\Delta_{pq}$ is largest for the edge connecting node 1 in layer 2 to node 1 in layer 1. That imbalance requires a net flow from node-layer pair $(1,2)$ to $(1,1)$. Since $a_1$ tunes the transition rate between layers \new{at node $n=1$, this net flow will} monotonically increase with $a_1$. Therefore, we expect the multiplex imbalance to be most sensitive to $a$  when $a_1$ changes with $a$. Parameter  $a_1$ is most sensitive to $a$ for the strategies of identical $a_n$ and increasing $a_n$ (i.e., $a_1=a$ in these cases), it randomly depends on $a$ for the  strategy of random $a_n$  (although it increases in expectation since  $\mathbb{E}[a_n] = (1+a)/2$), and it does not vary with $a$ for the strategy of decreasing $a_n$ (i.e., $a_1=1$). Therefore, our observed sensitivity with $a$ for the different strategies is \new{is exactly as one would} expect based on our understanding for how degree-imbalances affect   flow imbalances.

\new{\section{Application to brain-activity data}\label{sec:brain}} 

%
%
%

We  now 
study
 a MMC representation of a \new{functional}   brain network  \cite{guillon2017loss} with $N=148$ nodes (brain regions) and $I=7$ layers. \new{The data includes}  pairwise coherences of magnetoencephalography (MEG) signals  at  different  frequency ranges, $\{[1,4),[4,8),[8,10.5),[10.5,13),[13,20),[20,30),[30,45)\}$ (measured in Hz),  \new{and we interpret the matrices as intralayer adjacency matrices. We construct  intralayer transition matrices for them as described in Sec.~\ref{sec:app}.} 
 We uniformly couple the layers with an interlayer Markov chain with  transition matrix
\new{\begin{align}
\tilde{P}_{ij} = \left\{\begin{array}{rl}
 1,&  |i-j|=1\text{ and }  i\in\{1,7\} \\
  1/2,&  |i-j|=1\text{ and }  i\in\{2,\dots,6\} \\
  0,&  |i-j|\neq  1.
\end{array} \right.
\end{align}
In Appendix C, we   study node-specific transition matrices $\tilde{\bm{P}}^{(n)}$ that are similar to those described in Sec.~\ref{sec:optimal}.
}

%
%

\begin{figure}[h]
\vspace{.5cm}
\centering
\includegraphics[width=\linewidth]{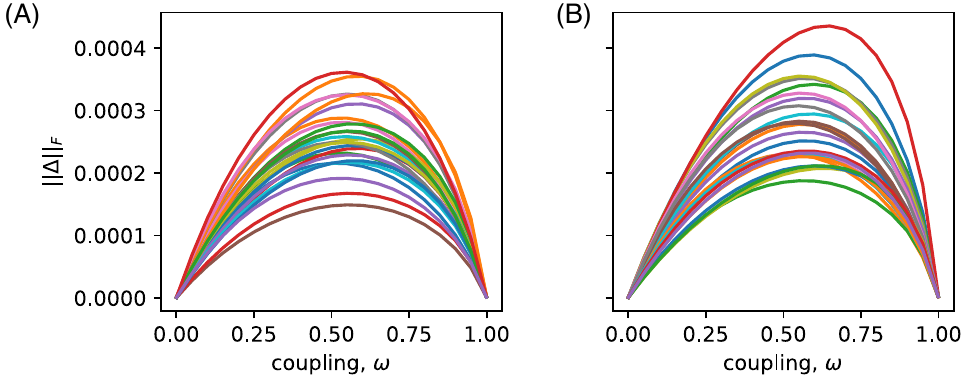}
\vspace{-.5cm}
\caption{
{\bf \new{Optimality} is different for persons with and without Alzheimers.}
We plot $||\Delta||_F$ versus $\omega$ for  25 persons (A) healthy persons
  and (B) persons with Alzheimer's disease. Observe that the $||\Delta ||_F$ values are slightly larger for persons with the disease, and the value of $\omega$ at which the optimum occurs, $\omega_\Delta^*$, shifts slightly to the right.
}
\label{fig:brain24}
\end{figure}

We  \new{first} conducted a population-level study of 
\new{the optimality of MMCs} for the 50 persons in the dataset   \cite{guillon2017loss}: 25  healthy persons and 25  persons with Alzheimer's disease. 
In  Fig.~\ref{fig:brain24}, we plot $||\Delta||_F$ versus $\omega$ for  (A)   healthy persons and (B) those with  Alzheimer's disease. Observe that the $||\Delta ||_F$ values are slightly larger for persons with the disease, and the value of $\omega$ at which the optimum occurs, $\omega_\Delta^*$, shifts slightly to the right. 
\new{Specifically}, when $\omega=0.5$, $||\Delta||_F$ is $6.6\%$ larger for persons with Alzheimers (0.000276 versus 0.000259). The average of $\omega_\Delta^* = \text{argmax}_\omega ||\Delta||_F$ was also found to be 2.1\% larger for persons with Alzheimers (0.570 versus 0.558). 
%


\new{Next, we    study the extent to which degree imbalance is a mechanism that helps drive the different  optimality of  MMCs for healthy and diseased brains. Recall our discussion in the last paragraph of Sec.~\ref{sec:convection} for how degree imbalance helped create the  convection cycle for the MMC that was shown in Fig.~\ref{fig:toy1}(B).}
%
\new{Now, we will show that a similar phenomenon occurs for the MMC representations of the brain data. 
That is,}   $\Delta_{pq}$ is often large for an interlayer edge associated with a node $n$ such that its intralayer degrees $d_{n}^{({i_p})} $ and $d_{n}^{({i_q})} $ are imbalanced.
%

\begin{figure*}[t]
\centering
\includegraphics[width= .65\linewidth]{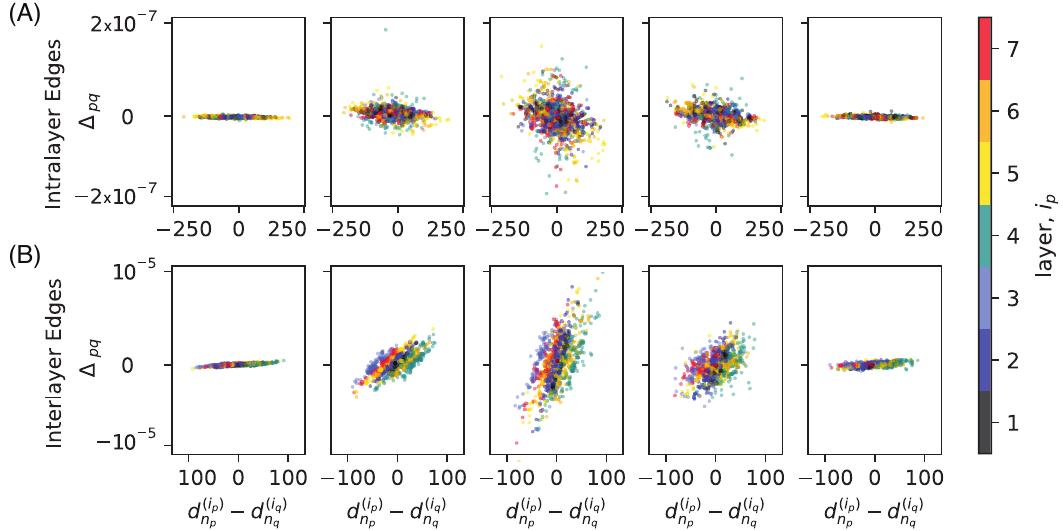}
\vspace{-.2cm}
\caption{
{\bf {Degree imbalance contributes to multiplex imbalance and convection} for  \new{an MMC representation of  brain-activity} \new{data}.}
We separately plot  $\Delta_{pq}$ versus $ (d_{n_p}^{(i_p)} - d_{n_q}^{(i_q)})$ for (A)~intralayer edges and (B) interlayer edges for a MMC representation of an empirical brain network with $N=148$ nodes (brain regions) and $I=7$ layers (correlated MEG signals at different  frequency ranges). 
From left to right, the different columns show results for different coupling strength:  $\omega\in\{0.01,0.1,0.5,0.9,0.99\}$.
Observe that $\Delta_{pq}$ and $(d_{n_p}^{(i_p)} - d_{n_q}^{(i_q)})$  are positively (negatively) correlated for interlayer (intralayer) edges, and $\Delta_{pq}$ have their largest magnitudes for $\omega=0.5$ (center column).
}
\label{fig:brain}
\end{figure*}

\begin{figure*}[t]
\centering
\includegraphics[width=.64\linewidth]{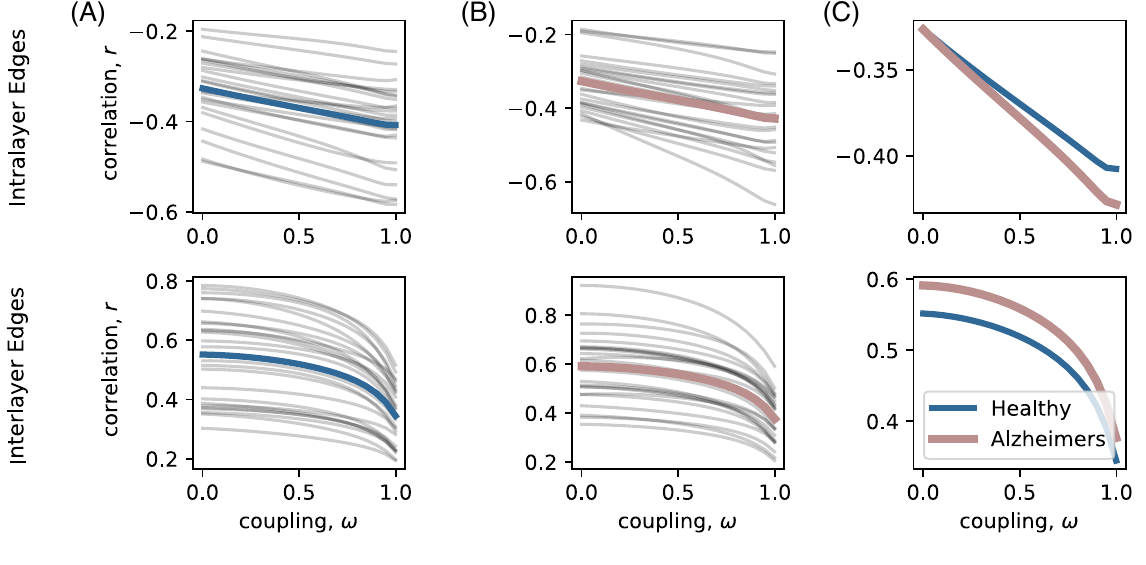}
\vspace{-.5cm}
\caption{
{\bf Correlation between multiplex imbalance and \new{intralayer-degree} imbalance.}
We plot the Pearson correlation coefficient \new{$r$} between $\Delta_{pq}$ and $d_{n_p}^{({i_p})} - d_{n_q}^{({j_q})}$ versus $\omega$ for (A) healthy persons   and (B) persons with Alzheimer's disease.  Different gray curves correspond to different people, and the thick colored curves indicate the mean curves. (C) A comparison of the mean curves for the two subpopulations.  All correlation coefficients are separately calculated  for the (top row)  intralayer edges and (bottom row) interlayer edges.
}
\vspace{-.1cm}
\label{fig:brain99}
\end{figure*}

In Fig.~\ref{fig:brain}, 
%
we plot  $\Delta_{pq}$   versus $(d_{n_p}^{({i_p})} - d_{n_q}^{({j_q})})$ separately for (A)~intralayer edges (i.e., $i_p$ = $i_q$) and (B) interlayer edges (i.e., $n_p$ = $n_q$).  
Different columns show results for different $\omega$.
First, observe   that   $\Delta_{pq}$  are approximately 50$\times$ larger for interlayer edges than  for intralayer edges [i.e., compare the y-axis of (B) to that of (A)].
Also, note that  $\Delta_{pq}$  obtain their largest magnitudes near $\omega=0.5$ (center column), which is consistent with our finding that $\omega_\Delta^*\approx 0.5$ for this MMC. 
\new{In Appendix C, we present extended results by studying the separate optimality of each $\Delta_{pq}$ versus $\omega$.}


The strong correlation between $\Delta_{pq}$ and $( d_{n}^{({i_p})} - d_{n}^{({j_q})})$  supports our hypothesis that the largest $\Delta_{pq}$ occur for interlayer edges associated with the largest intralayer-degree imbalance, $ (d_{n}^{({i_p})} - d_{n}^{({j_q})})$.
\new{We also find that this correlation differs between healthy and diseased brains.}
In   Fig.~\ref{fig:brain99}, we   plot the Pearson correlation coefficient between $\Delta_{pq}$ and $(d_{n_p}^{({i_p})} - d_{n_q}^{({j_q})})$ versus $\omega$ for  \new{(A) healthy persons and (B)  persons with Alzheimer's disease.} Different curves correspond to different people. The thick colored curves indicate \new{the subpopulations' mean values, and they are repeatedly shown in Fig.~\ref{fig:brain99}(C)} to highlight the difference between persons with/without the disease. We separately computed these correlation coefficients for intralayer and interlayer edges, which we show in the upper and lower rows, respectively.

%
%

Given our observation that imbalanced intralayer degrees contribute to multiplex imbalance and convection, and \new{that} both are increased for persons with Alzheimer's disease, our findings are consistent with previous work that found persons with Alzheimer's disease to have a loss of brain inter-frequency hubs \cite{guillon2017loss}.


\new{\section{Conclusion}\label{sec:conclusion}}

%
{Our work is motivated by interdisciplinary applications that use discrete Markov chains \cite{kendall1953stochastic,kingman1969markov,gilks1995markov,tierney1994markov,parr1998reinforcement} and by the observation that existing multiplex diffusion models \cite{gomez2013diffusion,sole2013spectral,radicchi2013abrupt,de2014navigability,tejedor2018diffusion,cencetti2019diffusive,trpevski2014discrete,dedom2015} are limited in their behavior \new{(see  Fig. 1)}. Here, we introduced a multiplex generalization of Markov chains that 
%
revealed  novel  phenomena: {multiplex convection} \new{and imbalance}. Convection cycles are a central topic in fluid mechanics, but they remain unexplored on networks. \new{We identified degree imbalances as one mechanism that contributes to convection, we showed that both  the extent of convection and the convergence rate are optimized   at intermediate coupling $\omega$. Finally, we developed an MMC-based study of frequency-multiplexed brain-activity data, finding that that the MMCs for persons with  Alzheimer's disease differ from those of healthy persons}.}
\new{Our work highlights MMCs and convection as two important new directions for network-science research.}

%
%
%

 ~

\begin{acknowledgments}
\new{We thank Per Sebastian Skardal,  Naoki Masuda and Sarah Muldoon, as well as the referees, for their helpful feedback. This work was supported in part by the  Simons Foundation (grant \#578333).}
\end{acknowledgments}

\appendix

\new{\section{Comparison to diffusions on  multiplex  networks}\label{sec:compare}}

The diffusion models that most closely resemble MMCs \new{are} the ones that \new{first define a \emph{supra-adjacency matrix}
\begin{align}
\new{\hat{\mathbb{A}}}(\omega)  = (1-\omega)  \textrm{diag}\left(\{{\bf A}^{(i)} \}\right) +  \omega   \new{\sum_n  \tilde{\bm{A}}^{(n)} \otimes  {\bf E}^{(n)}} 
\end{align}
that couples   intralayer adjacency matrices $\{{\bf A}^{(i)}\}$ with interlayer adjacency matrices $\{\tilde{\bm{ A}}^{(n)}\}$. Then one defines a transition matrix by neglecting that the  inter and intralayer edges are different,} 
\begin{align}
\hat{\mathbb{P}}(\omega) = \new{\hat{\mathbb{D}}}(\omega)^{-1}\new{\hat{\mathbb{A}}}(\omega),
\label{eq:supra_t2}
\end{align}
\new{where} $\new{\hat{\mathbb{D}}}(\omega)$ is a diagonal matrix  \new{in which the diagonal entries} encode the nodes' \emph{total degrees}, 
\begin{align}
[\new{\hat{\mathbb{D}}}(\omega)]_{pp}   =  \sum_q [\new{\hat{\mathbb{A}}}(\omega)]_{pq} = (1-\omega) d_{n_p}^{({i_p})} + \omega \tilde{d}_{i_p}^{(n_p)}.
\end{align}
\new{Equation}~\eqref{eq:supra_t2} is actually slightly different from the one defined in \new{\cite{mucha2010,sole2016random}}, but 
 theirs 
can be recovered by dividing $\new{\hat{\mathbb{A}}}(\omega)$ by $(1-\omega)$, so that their coupling strength is equivalent to $\omega/(1-\omega) \in[0,\infty)$. We use the definition of Eq.~\eqref{eq:supra_t2} since it allows us  to study the same range of $\omega$ as for MMCs, $\omega\in[0,1]$.
Also, it's worth noting that \cite{mucha2010} studied a continuous-time random walk with the goal of extending Markov stability \cite{delvenne2010stability}, whereas here we study a discrete-time random walk  {similar to \cite{sole2016random}}.
\new{We also note that these previous works focused on when the layers were uniformly coupled, $\tilde{\bm{A}}^{(n)} = \tilde{\bm{A}}~\forall n$.}

Because $\new{\hat{\mathbb{A}}}(\omega) $ is an adjacency matrix for an undirected network,    $\hat{\mathbb{P}}(\omega)$ has a stationary distribution with entries that are proportional to the degrees \cite{kemeny1976markov}---or in this case, the node-layer pairs: 
\begin{align}
\hat{\pi}_n^{(i)}(\omega) \varpropto (1-\omega) d_n^{(i)} + \omega \tilde{ d}_i^{(n)}.
\end{align} 
\new{Recall that $d_n^{(i)} = \sum_m A_{nm}^{({i})}$ and $\tilde{ d}_i^{(n)}  = \sum_j \tilde{A}_{ij}^{({n})}$ are the intralayer and interlayer degrees, respectively. In other words,
}
the $\hat{\pi}_n^{(i)}(\omega) $ values are not informative of a multiplex network's global (i.e., nonlocal) properties.  

~

\section{Perturbation Theory for Timescale Separation}\label{sec:weak}
%
\vspace{-.2cm}
We first study the dominant eigenvector of a supratransition matrix $\mathbb{P}(\omega)$ in the limit $\omega\to 0^+$, which corresponds to when transitions rarely occur using an interlayer Markov chain.

Let $\mu_1^{(i)}$, ${\bf v}^{(i)}$, and ${\bf u}^{(i)}$ denote the largest positive eigenvalue and corresponding left and right eigenvectors of each   transition matrix  ${\bf P}^{(i)}$ of the intralayer Markov chains, where $i\in\{1,\dots,I\}$ is a layer index. 
Furthermore, let $\tilde{\mu}_1^{(n)}$, $\tilde{\bm{ v}}^{(n)}$, and $\tilde{\bm{ u}}^{(n)}$ denote the same mathematical elements for each   transition matrix $\tilde{\bm{ P}}^{(n)}$ of the interlayer Markov chains, where $n\in\{1,\dots,N\}$ is a node indicx. 
Note that ${\mu}_1^{(i)}=\tilde{\mu}_1^{(n)}=1$ for any $i$ and $n$, since the transition matrices are row stochastic. Also, their corresponding right eigenvectors ${\bf u}^{(i)}=[1,\dots,1]^T$ and $\tilde{\bm{ u}}^{(n)}=[1,\dots,1]^T$ are vectors in which all the entries are ones. It is advantageous to let the left eigenvectors represent probability distributions, and so we normalize them in 1-norm.
We do not normalize the right eigenvectors (i.e., $\sum_n {  u}_n^{(i)} = N$ and $\sum_i \tilde{ { u}}^{(n)}_i=I$)  so that 
$ [{\bf v}^{(i)}]^T {\bf u}^{(i)} = \sum_n v^{(i)}_n= 1$ and $[\tilde{\bm{ v}}^{(n)}]^T \tilde{\bm{ u}}^{(n)} =\sum_i \tilde{v}^{(n)}_i=1$ for any $i$ and $n$.
Provided that the transition matrices ${\bf P}^{(i)}$ and $\tilde{\bm{ P}}^{(n)}$  are nonnegative and irreducible, Perron-Frobenius theory for nonnegative matrices \cite{bapat1997} guarantees that the left eigenvectors ${\bf{ v}}^{(i)}$ and $\tilde{\bm{ v}}^{(n)}$ are unique  and contain positive entries. 

Turning our attention to the spectra of  $\mathbb{P}(\omega)$, we denote its largest positive eigenvalue by $\lambda_1(\omega)$ and its   left and right eigenvectors by $\mathbbm{v} (\omega)$ and $\mathbbm{u}(\omega)$, respectively. %
We can write the dominant eigenvector equations as
\begin{align}\label{eq:evec}
\mathbb{P}(\omega)^T\mathbbm{v} (\omega) 
&=   \mathbbm{v} (\omega)\nonumber\\
\mathbb{P}(\omega) \mathbbm{u} (\omega)
&=   \mathbbm{u} (\omega) . 
\end{align}

Because $\mathbb{P}(\omega)$ is row stochastic for any $\omega\in[0,1]$, 
$\mathbbm{u}(\omega) =\mathbbm{u}    =   [1,\dots,1]^T$ is a right eigenvector with eigenvalue $\lambda_1(\omega) = \lambda_1 = 1$. That is, both are independent of $\omega$, and we can drop  $\omega$ as an argument. (This will be more rigorously supported in a theorem below.)

Provided that $\mathbb{P}(\omega)$ is a nonnegative irreducible matrix, $\mathbbm{v} (\omega)$ and $\mathbbm{u} (\omega)$ are uniquely defined and have positive entries \cite{bapat1997}. Note that this explicitly assumes $\omega\in(0,1)$. We denote the $\omega\to 0^+$ limits of  $\mathbbm{v} (\omega)$ and $\mathbbm{u} (\omega)$  by $\mathbbm{u} (0^+)$.
However, when $\omega=0$ (i.e., $\omega$ is exactly zero), then $\mathbb{P}(\omega)$ is not irreducible. We provide the following theorem to characterize the eigenspace associated with $\lambda_1=1$ in this case.


\begin{thm}\label{thm:uncoupled}
Let $\mathbb{P}(\omega)$ be a supracentrality matrix of a multiplex Markov chain and assume that each intralayer transition matrix ${\bf P}^{(i)}$ is nonnegative, irreducible.
Then the geometric and algebraic multiplicity of eigenvalue $\lambda_1=1$ of $\mathbb{P}(0)$ are both $I$ (recall that $I$ is the number of intralayer Markov chains), and the left and right eigenspaces are spanned by orthogonal eigenvectors 
\begin{align}
\mathbbm{v}^{(i)} &= {\bf e}^{(i)} \otimes  {\bf v}^{(i)}\nonumber\\ 
\mathbbm{u}^{(i)} &= {\bf e}^{(i)} \otimes  {\bf u}^{(i)},
\end{align}
 respectively, where ${\bf e}^{(i)}$ denotes the unit vector (i.e., all entries are zero except for the $i$-th entry, which is a $1$) and $\otimes$ denotes the Kronecker product. 
\end{thm}

\begin{remark}
We refer to the vectors $ \mathbbm{v}^{(i)}$ and $ \mathbbm{u}^{(i)}$ as  `block vectors', and they consist of zeros, except in the $i$-th blocks, which are ${\bf v}^{(i)}$ and ${\bf u}^{(i)}$, respectively.   
\end{remark}

\noindent{\bf Proof.} 
{\it 
First, we show that $\mathbbm{v}^{(i)}$ and $\mathbbm{u}^{(i)}$ are left and right eigenvectors of $\mathbb{P}(0)$ corresponding to the eigenvalue $\lambda_1=1$,
\begin{align}
\mathbb{P}(0) \mathbbm{u}^{(i)} 
	&= {\bf e}^{(i)} \otimes {\bf P}^{(i)}  {\bf u}^{(i)} \nonumber\\
	&= {\bf e}^{(i)} \otimes \mu_1^{(i)} 
	{\bf u}^{(i)}  \nonumber\\
	&= \mathbbm{u}^{(i)}
\end{align}
and
\begin{align}
\mathbb{P}(0)^T \mathbbm{v}^{(i)} 
	&= {\bf e}^{(i)} \otimes [{\bf P}^{(i)}]^T  {\bf v}^{(i)} \nonumber\\
	&= {\bf e}^{(i)} \otimes 
	\mu_1^{(i)}	 {\bf v}^{(i)}  \nonumber\\
	&= \mathbbm{v}^{(i)}.
\end{align}
It is also straightforward to show that these sets of eigenvectors are orthogonal:
\begin{align}
[\mathbbm{v}^{(i)}]^T\mathbbm{v}^{(j)} 
	&= ( {\bf e}^{(i)}  \otimes {\bf v}^{(i)}   )^T ( {\bf e}^{(j)} \otimes {\bf v}^{(j)} ) \nonumber\\
	&= ( [{\bf e}^{(i)}]^T  \otimes [{\bf v}^{(i)}   ]^T) ( {\bf e}^{(j)} \otimes {\bf v}^{(j)} ) \nonumber\\
	&=   [{\bf e}^{(i)}]^T {\bf e}^{(j)} \otimes [{\bf v}^{(i)}]^T {\bf v}^{(j)}  \nonumber\\
	&= \delta_{ij}  
\end{align}
and
\begin{align}
[\mathbbm{u}^{(i)}]^T\mathbbm{u}^{(j)} 
	&= ( {\bf e}^{(i)}  \otimes {\bf u}^{(i)}   )^T ( {\bf e}^{(j)} \otimes {\bf u}^{(j)} ) \nonumber\\
	&= ( [{\bf e}^{(i)}]^T  \otimes [{\bf u}^{(i)}   ]^T) ( {\bf e}^{(j)} \otimes {\bf u}^{(j)} ) \nonumber\\
	&=   [{\bf e}^{(i)}]^T {\bf e}^{(j)} \otimes [{\bf u}^{(i)}]^T {\bf u}^{(j)}  \nonumber\\
	&= \delta_{ij}    .
\end{align}
These results use that the Kronecker-product identity 
$(a\otimes b)(c\otimes d) = ac \otimes bd$ (assuming the dimensions appropriately match).

Provided that each ${\bf P}^{(i)}$ is nonnegative and irreducible, the dominant eigenvalue $\mu_1^{(i)}=1$ of each ${\bf P}^{(i)}$ has   geometric and algebraic multiplicity equal to 1. Thus the eigenvalue $\lambda_1=1$ of $\mathbb{P}(0)$ has geometric and algebraic multiplicity equal to $I$. 
The sets of eigenvectors $\{\mathbbm{v}^{(i)}\}$ and $\{\mathbbm{u}^{(i)}\}$ are eigenbases for the left and right eigenspaces for $\lambda_1=1$.
}

Next, we present our main analytical result for when there is a separation of time scales and transitions are far more likely to utilize an  intralayer Markov chain versus an  interlayer one.

\begin{thm} \label{thm:uncoupled2}
Let $\mathbb{P}(\omega)$ be a supracentrality matrix of a multiplex Markov chain and assume each intralayer transition matrix ${\bf P}^{(i)}$ is nonnegative, irreducible, and has a dominant eigenvalue such that $\mu_1^{(i)}=1$. 
We define $\mathbbm{v}(0^+)=\lim_{\omega \to 0^+} \mathbbm{v} (\omega)  $ 
as the limiting left 
  eigenvector of $\mathbb{P}(\omega)$. 
Then
\begin{align}\label{eq:lim_uv}
	  \mathbbm{v} (0^+) &= \sum_{i=1}^I   \tilde{\alpha}_{i} \mathbbm{v}^{(i)} , 
\end{align}
where the vector  $\tilde{\bm{\alpha}} = [\alpha_{1},\dots,\alpha_{I}]^T$ 
has positive entries that satisfy $\sum_{i=1}^I  \tilde{\alpha}_i $ 
and is a unique solution to
\begin{align}\label{eq:baah}
	 \tilde{\bm{ \alpha}}^T \tilde{\bm{X}}  &= 
	  \tilde{\bm{ \alpha}}^T. 
\end{align}
with
$	\tilde{{X}}_{ij}  =    \sum_{n=1}^N  \tilde{{P}}_{ij}^{(n)}  {v}_n^{(i)}. $
\end{thm}
 
 \begin{remark}
Since each $\tilde{\bm {P}}^{(n)}$ is row stochastic, matrix $\tilde{\bm{X}}$ is also row stochastic:
\begin{align}
\sum_j \tilde{{X}}_{ij} 
&=     \sum_{n=1}^N  \sum_j \tilde{{P}}_{ij}^{(n)}  {v}_n^{(i)}\nonumber\\
&=    \sum_{n=1}^N  {v}_n^{(i)}\nonumber\\
&=    1.
\end{align}
It follows that it  $ [1,\dots,1]^T$ is a right eigenvector of $\tilde{\bm{X}}$ for an eigenvalue equal to 1.
Therefore $\tilde{\bm{X}}$ is  an ``effective'' interlayer transition matrix that represents a type of weighted aggregation of the interlayer Markov chains $\{\bm{\tilde{P}}^{(n)}\}$.
\end{remark}

\begin{remark}
%
When the intralayer Markov chains are uniformly coupled, i.e., $\tilde{\bm{P}}^{(n)}=\tilde{\bm{P}}$ for each node $n$, it then follows   that $\tilde{\bm{X}}  = \tilde{\bm{P}}$ and
$\tilde{\bm{\alpha}} = \tilde{\bm{v}}$,
which is the left dominant eigenvector of $\tilde{\bm{P}}$.

\end{remark}

\begin{remark}
When the intralayer Markov chains have  doubly stochastic transition matrices, i.e., $\sum_n P_{nm}^{(i)}=\sum_m P_{nm}^{(i)}=1$ for each node $i$, 
then $v_n^{(i)}=1/N$ for each $i$ and
 $\tilde{\bm{X}} $ is    the mean intralayer transition matrix.
\end{remark} 
 

{\bf Proof.}
Theorem~\ref{thm:uncoupled} proved that $\lambda_{1}$ has a $P$-dimensional left   dominant eigenspace that are spanned by the left  eigenvectors $\mathbbm{v}^{(i)}$.
%
The continuity of eigenvector spaces \cite{kato2013perturbation} ensures that   $\mathbbm{v}(0^+)$
converges to lie within this subspace, which implies Eq.~\eqref{eq:lim_uv}. 
We now   prove that the constants $\tilde{\alpha}_i$ 
satisfy Eq.~\eqref{eq:baah}. 

We Taylor expand   $\mathbbm{v}(\omega)$ 
for small $\omega$     as
\begin{align}\label{eq:lddl}
	\mathbbm{v}(\omega) &=  \sum_{k=0}^K\omega^k\mathbbm{v}_k + \mathcal{O}(\omega^{K+1}) \,. 
\end{align}
Successive terms in this expansion represent higher-order derivatives of $\mathbbm{v}(\omega)$ 
with respect to $\omega$, and we assume that $\mathbbm{v}(\omega)$ has the appropriate smoothness [i.e., $\mathbbm{v}(\omega) \in C^{(k)}(0,1)$]. 
%
The $\omega \to  0^+$ limit of  $\mathbbm{v}(\omega)$    then becomes $\mathbbm{v}_0=\mathbbm{v}(0^+)$. 
Focusing on the first-order approximation, we insert $\mathbbm{v}(\omega) \approx \mathbbm{v}_0 + \omega \mathbbm{v}_1 $ 
into the eigenvalue equation
\begin{align}\label{eq:vv}
\mathbbm{v} (\omega) &=   \mathbb{P}(\omega)^T \mathbbm{v} (\omega)
\end{align}
to obtain
\begin{widetext}
\begin{align}\label{eq:vv}
\mathbbm{v}_0 + \omega \mathbbm{v}_1 
&= (1-\omega) \mathbb{P}(0)^T \left[ \mathbbm{v}_0 + \omega \mathbbm{v}_1 \right] 
+  \omega \sum_n  \left[(\tilde{\bm{P}}^{(n)} \otimes  {\bf E}^{(n)} \right]^T \left[ \mathbbm{v}_0 + \omega \mathbbm{v}_1 \right] \nonumber\\
&= \mathbb{P}(0)^T\mathbbm{v}_0 + \omega \mathbb{P}(0)^T \mathbbm{v}_1-  \omega \mathbb{P}(0)^T \mathbbm{v}_0  - \omega^2 \mathbb{P}(0)^T \mathbbm{v}_1 
+  \omega \sum_n  \left[(\tilde{\bm{P}}^{(n)} \otimes  {\bf E}^{(n)} \right]^T   \mathbbm{v}_0  
+  \omega^2 \sum_n  \left[(\tilde{\bm{P}}^{(n)} \otimes  {\bf E}^{(n)} \right]^T     \mathbbm{v}_1  
\end{align}
\end{widetext}
The second-order terms will be negligible as $\omega\to0$, and so we separately collect the zeroth-order and first-order terms in $\omega$ to obtain two consistency equations:
\begin{align}\label{eq:w0}
\mathbbm{v}_0  
&= \mathbb{P}(0)^T\mathbbm{v}_0 
\end{align}
and
\begin{align}\label{eq:w1}
  \mathbbm{v}_1 
&=     \mathbb{P}(0)^T \mathbbm{v}_1-    \mathbb{P}(0)^T \mathbbm{v}_0   +    \sum_n  \left[(\tilde{\bm{P}}^{(n)} \otimes  {\bf E}^{(n)} \right]^T   \mathbbm{v}_0  .
\end{align}
The consistency equation arising for the zeroth-order terms is exactly the eigenvalue equation with $\omega=0$, as expected. It implies a solution of the form given by Eq.~\eqref{eq:lim_uv}. 

To proceed, we
left multiply the consistency equation arising from the first-order terms by $\mathbbm{u}^{(i)}$, yielding 
\begin{widetext}
\begin{align}\label{eq:w1aa}
[\mathbbm{u}^{(i)}]^T \mathbbm{v}_1 
&=   [\mathbbm{u}^{(i)}]^T  \mathbb{P}(0)^T \mathbbm{v}_1-    [\mathbbm{u}^{(i)}]^T \mathbb{P}(0)^T \mathbbm{v}_0   
+    \sum_n  [\mathbbm{u}^{(i)}]^T (\tilde{\bm{P}}^{(n)} \otimes  {\bf E}^{(n)} )^T   \mathbbm{v}_0  .
\end{align}
\end{widetext}
However, $[\mathbbm{u}^{(i)}]^T \mathbb{P}(0)^T = [\mathbbm{u}^{(i)}]^T$ and the term on the left-hand side is canceled by the first term on the right-hand side, which yields
\begin{align}\label{eq:wd}
   [\mathbbm{u}^{(i)}]^T  \mathbbm{v}_0   =    \sum_n  \mathbbm{u}^{(i)} (\tilde{\bm{P}}^{(n)} \otimes  {\bf E}^{(n)} )^T   \mathbbm{v}_0  .
\end{align}

 \begin{figure*}[t]
\centering
\includegraphics[width=.6\linewidth]{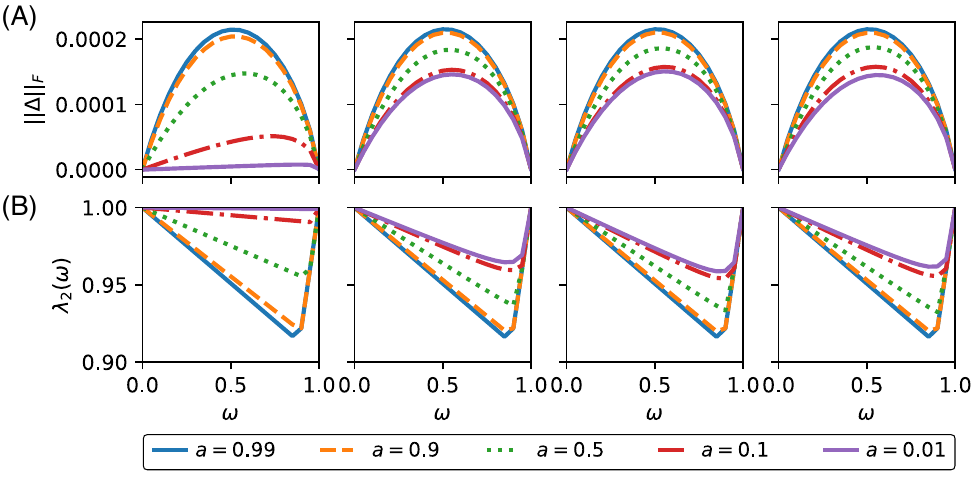}
\vspace{-.1cm}
\caption{
{\bf Optimality of multiplex imbalance and convergence rate for MMC model of a functional brain network}
(A) Total multiplex imbalance $||\Delta||_F$ and (B) convergence rate $\lambda_2(\omega)$  versus $\omega$ for the same intralayer Markov chains as in Figs.~5-7, except with node-specific interlayer Markov chains with  transition matrices having entries: $\tilde{P}^{(n)}_{ij} = (1-a_n)$ if $i=j$,  $\tilde{P}^{(n)}_{ij} = a_n$ if $i= |j\pm 1|$, and $\tilde{P}^{(n)}_{ij} = 0$ otherwise.
Each column depicts results for a different strategy for choosing the $a_n$ values, and each depends on a parameter $a\in[0,1]$: 
(left)~Identical $a_n$;
(center-left) Increasing $a_n$;  
(center-right) Decreasing $a_n$; and
(right) Random $a_n$.
}
\vspace{-.1cm}
\label{fig:brain007}
\end{figure*}

To simplify the left-hand side of Eq.~\eqref{eq:wd}, we use   
\begin{align}
[\mathbbm{u}^{(i)}]^T
&=[{\bf e}^{(i)} \otimes  {\bf u}^{(i)}]^T\nonumber\\
&=[{\bf e}^{(i)}]^T \otimes  [{\bf u}^{(i)}]^T
\end{align}
and $\mathbbm{v}_0 = \sum_j  \tilde{\alpha}_{j} \left( {\bf e}^{(j)} \otimes  {\bf v}^{(j)}\right)$ 
to obtain 
\begin{align}\label{eq:wdagd}
   [\mathbbm{u}^{(i)}]^T   \mathbbm{v}_0   
   &=    \sum_j  \tilde{\alpha}_{j} 
   ([{\bf e}^{(i)}]^T \otimes  [{\bf u}^{(i)}]^T ) 
   ({\bf e}^{(i)} \otimes  {\bf v}^{(j)}) \nonumber\\
   &=    \sum_j  \tilde{\alpha}_{j} 
   ([{\bf e}^{(i)}]^T    {\bf e}^{(j)} ) \otimes
   ( [{\bf u}^{(i)}]^T     {\bf v}^{(j)}) \nonumber\\
  &=  \sum_j \tilde{\alpha}_{j}  \delta_{ij}    ( [{\bf u}^{(i)}]^T     {\bf v}^{(j)}) \nonumber\\
  &= \tilde{\alpha}_{i}   .
\end{align}
(Recall that ${u}_n^{(i)}=1$ for each $n$ and $\sum_n {v}^{(i)}_n=1$.)
\new{Finally, we obtain
$	 \tilde{\bm{ \alpha}}^T \tilde{\bm{X}}   =
\tilde{\bm{ \alpha}}^T $
 by setting Eq.~\eqref{eq:wdagd} equal 
 to the following simplification for the}
right-hand side of Eq.~\eqref{eq:wd}, 
\begin{widetext}
\begin{align}\label{eq:rr}
\sum_n  [\mathbbm{u}^{(i)}]^T ( [ \tilde{\bm{P}}^{(n)}]^T \otimes  [{\bf E}^{(n)}]^T )  \mathbbm{v}_0
&=\sum_n    [\mathbbm{u}^{(i)}]^T( [ \tilde{\bm{P}}^{(n)}]^T \otimes  [{\bf E}^{(n)}]^T ) \sum_j \tilde{\alpha}_j \mathbbm{v}^{(j)}\nonumber\\
&=\sum_j \tilde{\alpha}_j  \sum_n    \left( [{\bf e}^{(i)}]^T \otimes  [{\bf u}^{(i)}]^T\right)( [ \tilde{\bm{P}}^{(n)}]^T \otimes  [{\bf E}^{(n)}]^T )  \left( {\bf e}^{(j)} \otimes  {\bf v}^{(j)}\right)\nonumber\\
&=\sum_j \tilde{\alpha}_j  \sum_n    \left( [{\bf e}^{(i)}]^T[ \tilde{\bm{P}}^{(n)}]^T  \otimes  [{\bf u}^{(i)}]^T [{\bf E}^{(n)}]^T \right) \left( {\bf e}^{(j)} \otimes  {\bf v}^{(j)}\right)\nonumber\\
&=\sum_j \tilde{\alpha}_j  \sum_n    \left( [{\bf e}^{(i)}]^T[ \tilde{\bm{P}}^{(n)}]^T{\bf e}^{(j)}  \right) \otimes   \left( [{\bf u}^{(i)}]^T [{\bf E}^{(n)}]^T {\bf v}^{(j)}\right)  \nonumber\\
&=\sum_j \tilde{\alpha}_j  \sum_n     \left[ \tilde{{P}}^{(n)}\right]_{ij}^T   {u}_n^{(i)}  {v}_n^{(j)}  \nonumber\\
&=\sum_j \tilde{\alpha}_j  \sum_n  
 \tilde{{P}}^{(n)}_{ji} {v}_n^{(j)}   \nonumber\\
 &=\sum_j \tilde{\alpha}_j  X_{ji}.
\end{align}
\end{widetext}
\vspace{-.2cm}


\section{Extended Study of MMC Model for\\ Frequency-Multiplexed Functional Brain Networks}

\vspace{-.2cm}
Here, we provide further results and insights on the optimality of multiplex imbalance for MMC models of  frequency-multiplexed functional brain networks using empirical data from \cite{guillon2017loss}. 
\new{Unlike our study in Sec.~\ref{sec:brain}, we now}
couple each node across layers using  node-specific interlayer Markov chains with  transition matrices  $\tilde{\bm{P}}^{(n)}$ having entries
\begin{equation}
\tilde{P}^{(n)}_{ij} = \left \{
\begin{array}{rl}
1-a_n,&i=j \\
 a_n,&|i-j| = 1 \text{ and } i\in\{1,N\} \\
 a_n/2,&|i-j| = 1 \text{ and } i\not\in\{1,N\} \\
0,&\text{otherwise}.
\end{array}\right. 
\end{equation}
That is, we couple the Markov chain layer for each frequency band with those of the adjacent frequency bands with weight $a_n/2$ (with the exception of the the first and last layers, which are coupled by  $a_n$).
\new{We choose the $a_n$ values in the same way as we described in Sec.~\ref{sec:optimal}. Because we would like to introduce correlations between the $a_n$ values and the nodes' intralayer degrees $d_n^{(i)}$, we permuted the nodes indices so that their mean intralayer degrees, $I^{-1}\sum_i d_n^{(i)}$, decrease monotonically with $n=1,2\dots, N$.
}

\begin{figure*}[t]
\centering
\includegraphics[width=.7\linewidth]{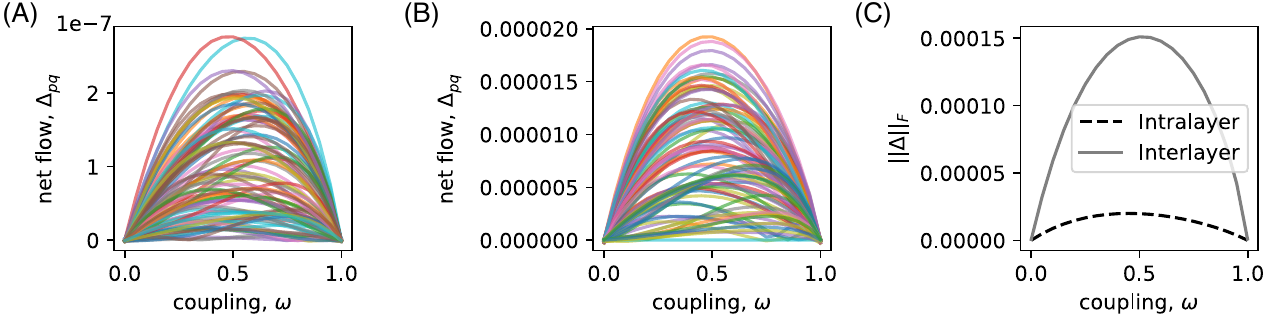}
\vspace{-.2cm}
\caption{
{\bf Optimality of $\Delta_{pq}$ for each edge \new{in the MMC representing  the brain data}.}
We plot the positive $\Delta_{pq}$ values \new{versus $\omega$} for each (A) intralayer and (B) interlayer edge. In panel (C), we   decompose $\Delta  = \Delta^{(intra)} + \Delta^{(inter)}$ into two matrices,  one corresponding to  intralayer edges and the other corresponding to interlayer edges, and we plot their respective Frobenius norms versus $\omega$.
}
\vspace{-.1cm}
\label{fig:brain3}
\end{figure*}

\begin{figure*}[t]
\centering
\includegraphics[width=.65\linewidth]{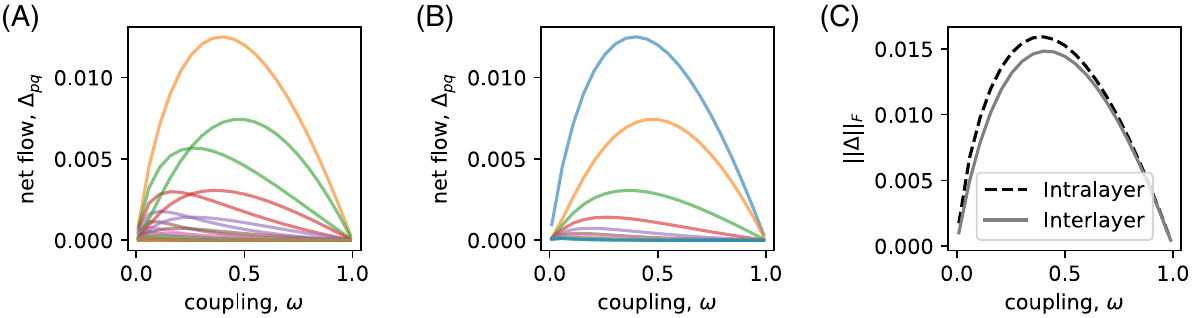}
\vspace{-.2cm}
\caption{
{\bf Optimality of $\Delta_{pq}$ for each  edge \new{for the MMC shown in Fig.~1}.}
\new{The results are similar to those  in Fig.~\ref{fig:brain3}. }
}
\label{fig:edge_opt}
\end{figure*}

In  Fig.~\ref{fig:brain007}, we  show a plot that is analogous to Fig.~3, except we now show results for the brain dataset.
The top panels depict  the total multiplex imbalance $||\Delta||_F$, while the lower  panels depict convergence rate $\lambda_2(\omega)$, both as a function of   $\omega$. 
Each column in Fig.~\ref{fig:brain007} depicts curves for a different strategy for choosing the $a_n$ values, and each depends on a parameter $a\in[0,1]$ (see the description of Fig.~3 in the main to remind yourself of these four strategies).
Observe that in this case, the locations $\omega_\Delta^*$ and $\omega_{\lambda_2}^*$ of optima occur at very different values of $\omega$, highlighting that optimality can more complicated when the number of Markov-chain layers increases (in this case there are $I=7$ layers, whereas Fig.~3 in the main text shows results for $I=2$ layers).

%
%
%
%
%
%
%
%
%
%
%

\new{To gain insight into how optimality may depend on the number of layers, next}
 we next study how \new{the net flow   $\Delta_{pq}$ across each edge obtains its own}   optimum at some value $\omega$.
%
In Fig.~\ref{fig:brain3}, we plot each $\Delta_{pq}$ versus $\omega$, and we separately plot  the values for (A)~intralayer edges [i.e., edges between node-layer pairs $(n_p,i_p)$ and $(n_q,i_q)$ with $i_q=i_p$] and (B)~interlayer edges [i.e., edges between node-layer pairs $(n_p,i_p)$ and $(n_q,i_q)$ with $n_q=n_p$]. 
That is, we create the decomposition  $\Delta  = \Delta^{(intra)} + \Delta^{(inter)}$.
Because   the values $\Delta_{pq} $ come in pairs having opposite signs, i.e., since $\Delta_{pq} =- \Delta_{qp}$, we only plot positive $\Delta_{pq}$.  Interesting, we find for all edges that the sign of $\Delta_{pq}$ does not change. That is, the directions of net flows do not switch as $\omega$ varies, although it remains unclear if this is a general property of  all multiplex networks.

Observe in Fig.~\ref{fig:brain3}(A) and (B) that the $\Delta_{pq}$ value for some edges become much larger than those of others, and  that the largest values obtain their maximum near $\omega\approx 0.5$. 
In  Fig.~\ref{fig:brain3}(C), we plot   $||\Delta^{(intra)}||_F$ and $||\Delta^{(inter)}||_F$, where $\Delta^{(intra)}_{pq}  = \Delta_{pq}$ if the edge between $p$ and $q$ is an intralayer edge and  $\Delta^{(intra)}_{pq}  = 0 $ otherwise.   $\Delta^{(inter)}_{pq}$ is similarly defined. Note that  $||\Delta ||_F^2 = ||\Delta^{(intra)}||_F^2 +  ||\Delta^{(inter)}||_F^2$. 
Observe that  the $\Delta_{pq}^{(intra)}$ values are on average about one-tenth as large as the $\Delta_{pq}^{(inter)}$ values.

\new{In 
Fig.~\ref{fig:edge_opt},
we show plots that are analogous to Fig.~\ref{fig:brain3}, except they are for the  MMC from Fig.~1}. 
%
Observe in Fig.~\ref{fig:edge_opt}(A) and (B) that the $\Delta_{pq}$ value for some edges become much larger than those of others, and  that the largest values obtain their maximum near $\omega\approx 0.45$. In contrast, the  $\Delta_{pq}$ values that never become  large tend to obtain their maximums at smaller $\omega \ll0.45$.  
\new{Observe in Fig.~\ref{fig:edge_opt}(C) that the  $\Delta_{pq}$ values for inter-- and intralayer edges for this 2-layer MMC are about the same magnitude. This contrasts Fig.~\ref{fig:brain3}(C) where the $\Delta_{pq}$ values are much larger for the interlayer edges.}

%
%
%
%
%

\bibliographystyle{apsrev4-1}
\bibliography{bibliography}

\begin{thebibliography}{43}%
\makeatletter
\providecommand \@ifxundefined [1]{%
 \@ifx{#1\undefined}
}%
\providecommand \@ifnum [1]{%
 \ifnum #1\expandafter \@firstoftwo
 \else \expandafter \@secondoftwo
 \fi
}%
\providecommand \@ifx [1]{%
 \ifx #1\expandafter \@firstoftwo
 \else \expandafter \@secondoftwo
 \fi
}%
\providecommand \natexlab [1]{#1}%
\providecommand \enquote  [1]{``#1''}%
\providecommand \bibnamefont  [1]{#1}%
\providecommand \bibfnamefont [1]{#1}%
\providecommand \citenamefont [1]{#1}%
\providecommand \href@noop [0]{\@secondoftwo}%
\providecommand \href [0]{\begingroup \@sanitize@url \@href}%
\providecommand \@href[1]{\@@startlink{#1}\@@href}%
\providecommand \@@href[1]{\endgroup#1\@@endlink}%
\providecommand \@sanitize@url [0]{\catcode `\\12\catcode `\$12\catcode
  `\&12\catcode `\#12\catcode `\^12\catcode `\_12\catcode `\%12\relax}%
\providecommand \@@startlink[1]{}%
\providecommand \@@endlink[0]{}%
\providecommand \url  [0]{\begingroup\@sanitize@url \@url }%
\providecommand \@url [1]{\endgroup\@href {#1}{\urlprefix }}%
\providecommand \urlprefix  [0]{URL }%
\providecommand \Eprint [0]{\href }%
\providecommand \doibase [0]{http://dx.doi.org/}%
\providecommand \selectlanguage [0]{\@gobble}%
\providecommand \bibinfo  [0]{\@secondoftwo}%
\providecommand \bibfield  [0]{\@secondoftwo}%
\providecommand \translation [1]{[#1]}%
\providecommand \BibitemOpen [0]{}%
\providecommand \bibitemStop [0]{}%
\providecommand \bibitemNoStop [0]{.\EOS\space}%
\providecommand \EOS [0]{\spacefactor3000\relax}%
\providecommand \BibitemShut  [1]{\csname bibitem#1\endcsname}%
\let\auto@bib@innerbib\@empty
\bibitem [{\citenamefont {Falkovich}(2011)}]{falkovich2011fluid}%
  \BibitemOpen
  \bibfield  {author} {\bibinfo {author} {\bibfnamefont {G.}~\bibnamefont
  {Falkovich}},\ }\href@noop {} {\emph {\bibinfo {title} {Fluid mechanics: A
  short course for physicists}}}\ (\bibinfo  {publisher} {Cambridge University
  Press},\ \bibinfo {year} {2011})\BibitemShut {NoStop}%
\bibitem [{\citenamefont {Cozzo}\ \emph {et~al.}(2018)\citenamefont {Cozzo},
  \citenamefont {De~Arruda}, \citenamefont {Rodrigues},\ and\ \citenamefont
  {Moreno}}]{cozzo2018multiplex}%
  \BibitemOpen
  \bibfield  {author} {\bibinfo {author} {\bibfnamefont {E.}~\bibnamefont
  {Cozzo}}, \bibinfo {author} {\bibfnamefont {G.~F.}\ \bibnamefont
  {De~Arruda}}, \bibinfo {author} {\bibfnamefont {F.~A.}\ \bibnamefont
  {Rodrigues}}, \ and\ \bibinfo {author} {\bibfnamefont {Y.}~\bibnamefont
  {Moreno}},\ }\href@noop {} {\emph {\bibinfo {title} {Multiplex Networks:
  Basic Formalism and Structural Properties}}}\ (\bibinfo  {publisher}
  {Springer},\ \bibinfo {year} {2018})\BibitemShut {NoStop}%
\bibitem [{\citenamefont {Boccaletti}\ \emph {et~al.}(2014)\citenamefont
  {Boccaletti}, \citenamefont {Bianconi}, \citenamefont {Criado}, \citenamefont
  {Del~Genio}, \citenamefont {G{\'o}mez-Garde{\~n}es}, \citenamefont {Romance},
  \citenamefont {Sendina-Nadal}, \citenamefont {Wang},\ and\ \citenamefont
  {Zanin}}]{boccaletti2014}%
  \BibitemOpen
  \bibfield  {author} {\bibinfo {author} {\bibfnamefont {S.}~\bibnamefont
  {Boccaletti}}, \bibinfo {author} {\bibfnamefont {G.}~\bibnamefont
  {Bianconi}}, \bibinfo {author} {\bibfnamefont {R.}~\bibnamefont {Criado}},
  \bibinfo {author} {\bibfnamefont {C.}~\bibnamefont {Del~Genio}}, \bibinfo
  {author} {\bibfnamefont {J.}~\bibnamefont {G{\'o}mez-Garde{\~n}es}}, \bibinfo
  {author} {\bibfnamefont {M.}~\bibnamefont {Romance}}, \bibinfo {author}
  {\bibfnamefont {I.}~\bibnamefont {Sendina-Nadal}}, \bibinfo {author}
  {\bibfnamefont {Z.}~\bibnamefont {Wang}}, \ and\ \bibinfo {author}
  {\bibfnamefont {M.}~\bibnamefont {Zanin}},\ }\href@noop {} {\bibfield
  {journal} {\bibinfo  {journal} {Physics Reports}\ }\textbf {\bibinfo {volume}
  {544}},\ \bibinfo {pages} {1} (\bibinfo {year} {2014})}\BibitemShut {NoStop}%
\bibitem [{\citenamefont {Kivel{\"a}}\ \emph {et~al.}(2014)\citenamefont
  {Kivel{\"a}}, \citenamefont {Arenas}, \citenamefont {Barthelemy},
  \citenamefont {Gleeson}, \citenamefont {Moreno},\ and\ \citenamefont
  {Porter}}]{kivela2014}%
  \BibitemOpen
  \bibfield  {author} {\bibinfo {author} {\bibfnamefont {M.}~\bibnamefont
  {Kivel{\"a}}}, \bibinfo {author} {\bibfnamefont {A.}~\bibnamefont {Arenas}},
  \bibinfo {author} {\bibfnamefont {M.}~\bibnamefont {Barthelemy}}, \bibinfo
  {author} {\bibfnamefont {J.~P.}\ \bibnamefont {Gleeson}}, \bibinfo {author}
  {\bibfnamefont {Y.}~\bibnamefont {Moreno}}, \ and\ \bibinfo {author}
  {\bibfnamefont {M.~A.}\ \bibnamefont {Porter}},\ }\href@noop {} {\bibfield
  {journal} {\bibinfo  {journal} {Journal of Complex Networks}\ }\textbf
  {\bibinfo {volume} {2}},\ \bibinfo {pages} {203} (\bibinfo {year}
  {2014})}\BibitemShut {NoStop}%
\bibitem [{\citenamefont {Taylor}\ \emph {et~al.}(2015)\citenamefont {Taylor},
  \citenamefont {Klimm}, \citenamefont {Harrington}, \citenamefont
  {Kram{\'a}r}, \citenamefont {Mischaikow}, \citenamefont {Porter},\ and\
  \citenamefont {Mucha}}]{taylor2015topological}%
  \BibitemOpen
  \bibfield  {author} {\bibinfo {author} {\bibfnamefont {D.}~\bibnamefont
  {Taylor}}, \bibinfo {author} {\bibfnamefont {F.}~\bibnamefont {Klimm}},
  \bibinfo {author} {\bibfnamefont {H.~A.}\ \bibnamefont {Harrington}},
  \bibinfo {author} {\bibfnamefont {M.}~\bibnamefont {Kram{\'a}r}}, \bibinfo
  {author} {\bibfnamefont {K.}~\bibnamefont {Mischaikow}}, \bibinfo {author}
  {\bibfnamefont {M.~A.}\ \bibnamefont {Porter}}, \ and\ \bibinfo {author}
  {\bibfnamefont {P.~J.}\ \bibnamefont {Mucha}},\ }\href@noop {} {\bibfield
  {journal} {\bibinfo  {journal} {Nature Communications}\ }\textbf {\bibinfo
  {volume} {6}},\ \bibinfo {pages} {7723} (\bibinfo {year} {2015})}\BibitemShut
  {NoStop}%
\bibitem [{\citenamefont {Strano}\ \emph {et~al.}(2015)\citenamefont {Strano},
  \citenamefont {Shai}, \citenamefont {Dobson},\ and\ \citenamefont
  {Barthelemy}}]{strano2015multiplex}%
  \BibitemOpen
  \bibfield  {author} {\bibinfo {author} {\bibfnamefont {E.}~\bibnamefont
  {Strano}}, \bibinfo {author} {\bibfnamefont {S.}~\bibnamefont {Shai}},
  \bibinfo {author} {\bibfnamefont {S.}~\bibnamefont {Dobson}}, \ and\ \bibinfo
  {author} {\bibfnamefont {M.}~\bibnamefont {Barthelemy}},\ }\href@noop {}
  {\bibfield  {journal} {\bibinfo  {journal} {Journal of The Royal Society
  Interface}\ }\textbf {\bibinfo {volume} {12}},\ \bibinfo {pages} {20150651}
  (\bibinfo {year} {2015})}\BibitemShut {NoStop}%
\bibitem [{\citenamefont {Sol{\'e}-Ribalta}\ \emph {et~al.}(2019)\citenamefont
  {Sol{\'e}-Ribalta}, \citenamefont {Arenas},\ and\ \citenamefont
  {G{\'o}mez}}]{sole2019effect}%
  \BibitemOpen
  \bibfield  {author} {\bibinfo {author} {\bibfnamefont {A.}~\bibnamefont
  {Sol{\'e}-Ribalta}}, \bibinfo {author} {\bibfnamefont {A.}~\bibnamefont
  {Arenas}}, \ and\ \bibinfo {author} {\bibfnamefont {S.}~\bibnamefont
  {G{\'o}mez}},\ }\href@noop {} {\bibfield  {journal} {\bibinfo  {journal} {New
  Journal of Physics}\ }\textbf {\bibinfo {volume} {21}},\ \bibinfo {pages}
  {035003} (\bibinfo {year} {2019})}\BibitemShut {NoStop}%
\bibitem [{\citenamefont {Haimes}\ and\ \citenamefont
  {Jiang}(2001)}]{haimes2001leontief}%
  \BibitemOpen
  \bibfield  {author} {\bibinfo {author} {\bibfnamefont {Y.~Y.}\ \bibnamefont
  {Haimes}}\ and\ \bibinfo {author} {\bibfnamefont {P.}~\bibnamefont {Jiang}},\
  }\href@noop {} {\bibfield  {journal} {\bibinfo  {journal} {Journal of
  Infrastructure Systems}\ }\textbf {\bibinfo {volume} {7}},\ \bibinfo {pages}
  {1} (\bibinfo {year} {2001})}\BibitemShut {NoStop}%
\bibitem [{\citenamefont {Krackhardt}(1987)}]{krackhardt1987cognitive}%
  \BibitemOpen
  \bibfield  {author} {\bibinfo {author} {\bibfnamefont {D.}~\bibnamefont
  {Krackhardt}},\ }\href@noop {} {\bibfield  {journal} {\bibinfo  {journal}
  {Social Networks}\ }\textbf {\bibinfo {volume} {9}},\ \bibinfo {pages} {109}
  (\bibinfo {year} {1987})}\BibitemShut {NoStop}%
\bibitem [{\citenamefont {De~Domenico}\ \emph
  {et~al.}(2015{\natexlab{a}})\citenamefont {De~Domenico}, \citenamefont
  {Nicosia}, \citenamefont {Arenas},\ and\ \citenamefont
  {Latora}}]{de2015structural}%
  \BibitemOpen
  \bibfield  {author} {\bibinfo {author} {\bibfnamefont {M.}~\bibnamefont
  {De~Domenico}}, \bibinfo {author} {\bibfnamefont {V.}~\bibnamefont
  {Nicosia}}, \bibinfo {author} {\bibfnamefont {A.}~\bibnamefont {Arenas}}, \
  and\ \bibinfo {author} {\bibfnamefont {V.}~\bibnamefont {Latora}},\
  }\href@noop {} {\bibfield  {journal} {\bibinfo  {journal} {Nature
  Communications}\ }\textbf {\bibinfo {volume} {6}},\ \bibinfo {pages} {1}
  (\bibinfo {year} {2015}{\natexlab{a}})}\BibitemShut {NoStop}%
\bibitem [{\citenamefont {Taylor}\ \emph
  {et~al.}(2016{\natexlab{a}})\citenamefont {Taylor}, \citenamefont {Shai},
  \citenamefont {Stanley},\ and\ \citenamefont {Mucha}}]{taylor2016enhanced}%
  \BibitemOpen
  \bibfield  {author} {\bibinfo {author} {\bibfnamefont {D.}~\bibnamefont
  {Taylor}}, \bibinfo {author} {\bibfnamefont {S.}~\bibnamefont {Shai}},
  \bibinfo {author} {\bibfnamefont {N.}~\bibnamefont {Stanley}}, \ and\
  \bibinfo {author} {\bibfnamefont {P.~J.}\ \bibnamefont {Mucha}},\ }\href@noop
  {} {\bibfield  {journal} {\bibinfo  {journal} {Physical Review Letters}\
  }\textbf {\bibinfo {volume} {116}},\ \bibinfo {pages} {228301} (\bibinfo
  {year} {2016}{\natexlab{a}})}\BibitemShut {NoStop}%
\bibitem [{\citenamefont {Taylor}\ \emph
  {et~al.}(2017{\natexlab{a}})\citenamefont {Taylor}, \citenamefont {Caceres},\
  and\ \citenamefont {Mucha}}]{taylor2016_b}%
  \BibitemOpen
  \bibfield  {author} {\bibinfo {author} {\bibfnamefont {D.}~\bibnamefont
  {Taylor}}, \bibinfo {author} {\bibfnamefont {R.~S.}\ \bibnamefont {Caceres}},
  \ and\ \bibinfo {author} {\bibfnamefont {P.~J.}\ \bibnamefont {Mucha}},\
  }\href@noop {} {\bibfield  {journal} {\bibinfo  {journal} {Physical Review
  X}\ }\textbf {\bibinfo {volume} {7}},\ \bibinfo {pages} {031056} (\bibinfo
  {year} {2017}{\natexlab{a}})}\BibitemShut {NoStop}%
\bibitem [{\citenamefont {Guillon}\ \emph {et~al.}(2017)\citenamefont
  {Guillon}, \citenamefont {Attal}, \citenamefont {Colliot}, \citenamefont
  {La~Corte}, \citenamefont {Dubois}, \citenamefont {Schwartz}, \citenamefont
  {Chavez},\ and\ \citenamefont {Fallani}}]{guillon2017loss}%
  \BibitemOpen
  \bibfield  {author} {\bibinfo {author} {\bibfnamefont {J.}~\bibnamefont
  {Guillon}}, \bibinfo {author} {\bibfnamefont {Y.}~\bibnamefont {Attal}},
  \bibinfo {author} {\bibfnamefont {O.}~\bibnamefont {Colliot}}, \bibinfo
  {author} {\bibfnamefont {V.}~\bibnamefont {La~Corte}}, \bibinfo {author}
  {\bibfnamefont {B.}~\bibnamefont {Dubois}}, \bibinfo {author} {\bibfnamefont
  {D.}~\bibnamefont {Schwartz}}, \bibinfo {author} {\bibfnamefont
  {M.}~\bibnamefont {Chavez}}, \ and\ \bibinfo {author} {\bibfnamefont
  {F.~D.~V.}\ \bibnamefont {Fallani}},\ }\href@noop {} {\bibfield  {journal}
  {\bibinfo  {journal} {Scientific Reports}\ }\textbf {\bibinfo {volume} {7}},\
  \bibinfo {pages} {1} (\bibinfo {year} {2017})}\BibitemShut {NoStop}%
\bibitem [{\citenamefont {Soriano-Pa{\~n}os}\ \emph {et~al.}(2018)\citenamefont
  {Soriano-Pa{\~n}os}, \citenamefont {Lotero}, \citenamefont {Arenas},\ and\
  \citenamefont {G{\'o}mez-Garde{\~n}es}}]{soriano2018spreading}%
  \BibitemOpen
  \bibfield  {author} {\bibinfo {author} {\bibfnamefont {D.}~\bibnamefont
  {Soriano-Pa{\~n}os}}, \bibinfo {author} {\bibfnamefont {L.}~\bibnamefont
  {Lotero}}, \bibinfo {author} {\bibfnamefont {A.}~\bibnamefont {Arenas}}, \
  and\ \bibinfo {author} {\bibfnamefont {J.}~\bibnamefont
  {G{\'o}mez-Garde{\~n}es}},\ }\href@noop {} {\bibfield  {journal} {\bibinfo
  {journal} {Physical Review X}\ }\textbf {\bibinfo {volume} {8}},\ \bibinfo
  {pages} {031039} (\bibinfo {year} {2018})}\BibitemShut {NoStop}%
\bibitem [{\citenamefont {Kemeny}\ and\ \citenamefont
  {Snell}(1976)}]{kemeny1976markov}%
  \BibitemOpen
  \bibfield  {author} {\bibinfo {author} {\bibfnamefont {J.~G.}\ \bibnamefont
  {Kemeny}}\ and\ \bibinfo {author} {\bibfnamefont {J.~L.}\ \bibnamefont
  {Snell}},\ }\href@noop {} {\emph {\bibinfo {title} {Markov Chains}}}\
  (\bibinfo  {publisher} {Springer-Verlag, New York},\ \bibinfo {year}
  {1976})\BibitemShut {NoStop}%
\bibitem [{\citenamefont {Kendall}(1953)}]{kendall1953stochastic}%
  \BibitemOpen
  \bibfield  {author} {\bibinfo {author} {\bibfnamefont {D.~G.}\ \bibnamefont
  {Kendall}},\ }\href@noop {} {\bibfield  {journal} {\bibinfo  {journal} {The
  Annals of Mathematical Statistics}\ }\textbf {\bibinfo {volume} {24}},\
  \bibinfo {pages} {338} (\bibinfo {year} {1953})}\BibitemShut {NoStop}%
\bibitem [{\citenamefont {Kingman}(1969)}]{kingman1969markov}%
  \BibitemOpen
  \bibfield  {author} {\bibinfo {author} {\bibfnamefont {J.}~\bibnamefont
  {Kingman}},\ }\href@noop {} {\bibfield  {journal} {\bibinfo  {journal}
  {Journal of Applied Probability}\ }\textbf {\bibinfo {volume} {6}},\ \bibinfo
  {pages} {1} (\bibinfo {year} {1969})}\BibitemShut {NoStop}%
\bibitem [{\citenamefont {Gilks}\ \emph {et~al.}(1995)\citenamefont {Gilks},
  \citenamefont {Richardson},\ and\ \citenamefont
  {Spiegelhalter}}]{gilks1995markov}%
  \BibitemOpen
  \bibfield  {author} {\bibinfo {author} {\bibfnamefont {W.~R.}\ \bibnamefont
  {Gilks}}, \bibinfo {author} {\bibfnamefont {S.}~\bibnamefont {Richardson}}, \
  and\ \bibinfo {author} {\bibfnamefont {D.}~\bibnamefont {Spiegelhalter}},\
  }\href@noop {} {\emph {\bibinfo {title} {Markov Chain Monte Carlo in
  Practice}}}\ (\bibinfo  {publisher} {Chapman and Hall/CRC},\ \bibinfo {year}
  {1995})\BibitemShut {NoStop}%
\bibitem [{\citenamefont {Tierney}(1994)}]{tierney1994markov}%
  \BibitemOpen
  \bibfield  {author} {\bibinfo {author} {\bibfnamefont {L.}~\bibnamefont
  {Tierney}},\ }\href@noop {} {\bibfield  {journal} {\bibinfo  {journal} {The
  Annals of Statistics}\ ,\ \bibinfo {pages} {1701}} (\bibinfo {year}
  {1994})}\BibitemShut {NoStop}%
\bibitem [{\citenamefont {Parr}\ and\ \citenamefont
  {Russell}(1998)}]{parr1998reinforcement}%
  \BibitemOpen
  \bibfield  {author} {\bibinfo {author} {\bibfnamefont {R.}~\bibnamefont
  {Parr}}\ and\ \bibinfo {author} {\bibfnamefont {S.~J.}\ \bibnamefont
  {Russell}},\ }in\ \href@noop {} {\emph {\bibinfo {booktitle} {Advances in
  Neural Information Processing Systems}}}\ (\bibinfo {year} {1998})\ pp.\
  \bibinfo {pages} {1043--1049}\BibitemShut {NoStop}%
\bibitem [{\citenamefont {Delvenne}\ \emph {et~al.}(2010)\citenamefont
  {Delvenne}, \citenamefont {Yaliraki},\ and\ \citenamefont
  {Barahona}}]{delvenne2010stability}%
  \BibitemOpen
  \bibfield  {author} {\bibinfo {author} {\bibfnamefont {J.-C.}\ \bibnamefont
  {Delvenne}}, \bibinfo {author} {\bibfnamefont {S.~N.}\ \bibnamefont
  {Yaliraki}}, \ and\ \bibinfo {author} {\bibfnamefont {M.}~\bibnamefont
  {Barahona}},\ }\href@noop {} {\bibfield  {journal} {\bibinfo  {journal}
  {Proceedings of the National Academy of Sciences}\ }\textbf {\bibinfo
  {volume} {107}},\ \bibinfo {pages} {12755} (\bibinfo {year}
  {2010})}\BibitemShut {NoStop}%
\bibitem [{\citenamefont {Schaub}\ \emph {et~al.}(2012)\citenamefont {Schaub},
  \citenamefont {Delvenne}, \citenamefont {Yaliraki},\ and\ \citenamefont
  {Barahona}}]{schaub2012markov}%
  \BibitemOpen
  \bibfield  {author} {\bibinfo {author} {\bibfnamefont {M.~T.}\ \bibnamefont
  {Schaub}}, \bibinfo {author} {\bibfnamefont {J.-C.}\ \bibnamefont
  {Delvenne}}, \bibinfo {author} {\bibfnamefont {S.~N.}\ \bibnamefont
  {Yaliraki}}, \ and\ \bibinfo {author} {\bibfnamefont {M.}~\bibnamefont
  {Barahona}},\ }\href@noop {} {\bibfield  {journal} {\bibinfo  {journal} {PloS
  one}\ }\textbf {\bibinfo {volume} {7}},\ \bibinfo {pages} {e32210} (\bibinfo
  {year} {2012})}\BibitemShut {NoStop}%
\bibitem [{\citenamefont {Mucha}\ \emph {et~al.}(2010)\citenamefont {Mucha},
  \citenamefont {Richardson}, \citenamefont {Macon}, \citenamefont {Porter},\
  and\ \citenamefont {Onnela}}]{mucha2010}%
  \BibitemOpen
  \bibfield  {author} {\bibinfo {author} {\bibfnamefont {P.~J.}\ \bibnamefont
  {Mucha}}, \bibinfo {author} {\bibfnamefont {T.}~\bibnamefont {Richardson}},
  \bibinfo {author} {\bibfnamefont {K.}~\bibnamefont {Macon}}, \bibinfo
  {author} {\bibfnamefont {M.~A.}\ \bibnamefont {Porter}}, \ and\ \bibinfo
  {author} {\bibfnamefont {J.-P.}\ \bibnamefont {Onnela}},\ }\href {\doibase
  10.1126/science.1184819} {\bibfield  {journal} {\bibinfo  {journal}
  {Science}\ }\textbf {\bibinfo {volume} {328}},\ \bibinfo {pages} {876}
  (\bibinfo {year} {2010})}\BibitemShut {NoStop}%
\bibitem [{\citenamefont {G\'{o}mez}\ \emph {et~al.}(2013)\citenamefont
  {G\'{o}mez}, \citenamefont {D\'{i}az-Guilera}, \citenamefont
  {G\'{o}mez-Garde\~{n}es}, \citenamefont {P\'{e}rez-Vicente}, \citenamefont
  {Moreno},\ and\ \citenamefont {Arenas}}]{gomez2013diffusion}%
  \BibitemOpen
  \bibfield  {author} {\bibinfo {author} {\bibfnamefont {S.}~\bibnamefont
  {G\'{o}mez}}, \bibinfo {author} {\bibfnamefont {A.}~\bibnamefont
  {D\'{i}az-Guilera}}, \bibinfo {author} {\bibfnamefont {J.}~\bibnamefont
  {G\'{o}mez-Garde\~{n}es}}, \bibinfo {author} {\bibfnamefont {C.~J.}\
  \bibnamefont {P\'{e}rez-Vicente}}, \bibinfo {author} {\bibfnamefont
  {Y.}~\bibnamefont {Moreno}}, \ and\ \bibinfo {author} {\bibfnamefont
  {A.}~\bibnamefont {Arenas}},\ }\href@noop {} {\bibfield  {journal} {\bibinfo
  {journal} {Physical Review Letters}\ }\textbf {\bibinfo {volume} {110}},\
  \bibinfo {pages} {028701} (\bibinfo {year} {2013})}\BibitemShut {NoStop}%
\bibitem [{\citenamefont {Sol\'{e}-Ribalta}\ \emph {et~al.}(2013)\citenamefont
  {Sol\'{e}-Ribalta}, \citenamefont {De~Domenico}, \citenamefont {Kouvaris},
  \citenamefont {Diaz-Guilera}, \citenamefont {Gomez},\ and\ \citenamefont
  {Arenas}}]{sole2013spectral}%
  \BibitemOpen
  \bibfield  {author} {\bibinfo {author} {\bibfnamefont {A.}~\bibnamefont
  {Sol\'{e}-Ribalta}}, \bibinfo {author} {\bibfnamefont {M.}~\bibnamefont
  {De~Domenico}}, \bibinfo {author} {\bibfnamefont {N.~E.}\ \bibnamefont
  {Kouvaris}}, \bibinfo {author} {\bibfnamefont {A.}~\bibnamefont
  {Diaz-Guilera}}, \bibinfo {author} {\bibfnamefont {S.}~\bibnamefont {Gomez}},
  \ and\ \bibinfo {author} {\bibfnamefont {A.}~\bibnamefont {Arenas}},\
  }\href@noop {} {\bibfield  {journal} {\bibinfo  {journal} {Physical Review
  E}\ }\textbf {\bibinfo {volume} {88}},\ \bibinfo {pages} {032807} (\bibinfo
  {year} {2013})}\BibitemShut {NoStop}%
\bibitem [{\citenamefont {Radicchi}\ and\ \citenamefont
  {Arenas}(2013)}]{radicchi2013abrupt}%
  \BibitemOpen
  \bibfield  {author} {\bibinfo {author} {\bibfnamefont {F.}~\bibnamefont
  {Radicchi}}\ and\ \bibinfo {author} {\bibfnamefont {A.}~\bibnamefont
  {Arenas}},\ }\href@noop {} {\bibfield  {journal} {\bibinfo  {journal} {Nature
  Physics}\ }\textbf {\bibinfo {volume} {9}},\ \bibinfo {pages} {717} (\bibinfo
  {year} {2013})}\BibitemShut {NoStop}%
\bibitem [{\citenamefont {De~Domenico}\ \emph {et~al.}(2014)\citenamefont
  {De~Domenico}, \citenamefont {Sol{\'e}-Ribalta}, \citenamefont {G{\'o}mez},\
  and\ \citenamefont {Arenas}}]{de2014navigability}%
  \BibitemOpen
  \bibfield  {author} {\bibinfo {author} {\bibfnamefont {M.}~\bibnamefont
  {De~Domenico}}, \bibinfo {author} {\bibfnamefont {A.}~\bibnamefont
  {Sol{\'e}-Ribalta}}, \bibinfo {author} {\bibfnamefont {S.}~\bibnamefont
  {G{\'o}mez}}, \ and\ \bibinfo {author} {\bibfnamefont {A.}~\bibnamefont
  {Arenas}},\ }\href@noop {} {\bibfield  {journal} {\bibinfo  {journal}
  {Proceedings of the National Academy of Sciences}\ }\textbf {\bibinfo
  {volume} {111}},\ \bibinfo {pages} {8351} (\bibinfo {year}
  {2014})}\BibitemShut {NoStop}%
\bibitem [{\citenamefont {Tejedor}\ \emph {et~al.}(2018)\citenamefont
  {Tejedor}, \citenamefont {Longjas}, \citenamefont {Foufoula-Georgiou},
  \citenamefont {Georgiou},\ and\ \citenamefont
  {Moreno}}]{tejedor2018diffusion}%
  \BibitemOpen
  \bibfield  {author} {\bibinfo {author} {\bibfnamefont {A.}~\bibnamefont
  {Tejedor}}, \bibinfo {author} {\bibfnamefont {A.}~\bibnamefont {Longjas}},
  \bibinfo {author} {\bibfnamefont {E.}~\bibnamefont {Foufoula-Georgiou}},
  \bibinfo {author} {\bibfnamefont {T.~T.}\ \bibnamefont {Georgiou}}, \ and\
  \bibinfo {author} {\bibfnamefont {Y.}~\bibnamefont {Moreno}},\ }\href@noop {}
  {\bibfield  {journal} {\bibinfo  {journal} {Physical Review X}\ }\textbf
  {\bibinfo {volume} {8}},\ \bibinfo {pages} {031071} (\bibinfo {year}
  {2018})}\BibitemShut {NoStop}%
\bibitem [{\citenamefont {Cencetti}\ and\ \citenamefont
  {Battiston}(2019)}]{cencetti2019diffusive}%
  \BibitemOpen
  \bibfield  {author} {\bibinfo {author} {\bibfnamefont {G.}~\bibnamefont
  {Cencetti}}\ and\ \bibinfo {author} {\bibfnamefont {F.}~\bibnamefont
  {Battiston}},\ }\href@noop {} {\bibfield  {journal} {\bibinfo  {journal} {New
  Journal of Physics}\ }\textbf {\bibinfo {volume} {21}},\ \bibinfo {pages}
  {035006} (\bibinfo {year} {2019})}\BibitemShut {NoStop}%
\bibitem [{\citenamefont {Trpevski}\ \emph {et~al.}(2014)\citenamefont
  {Trpevski}, \citenamefont {Stanoev}, \citenamefont {Koseska},\ and\
  \citenamefont {Kocarev}}]{trpevski2014discrete}%
  \BibitemOpen
  \bibfield  {author} {\bibinfo {author} {\bibfnamefont {I.}~\bibnamefont
  {Trpevski}}, \bibinfo {author} {\bibfnamefont {A.}~\bibnamefont {Stanoev}},
  \bibinfo {author} {\bibfnamefont {A.}~\bibnamefont {Koseska}}, \ and\
  \bibinfo {author} {\bibfnamefont {L.}~\bibnamefont {Kocarev}},\ }\href@noop
  {} {\bibfield  {journal} {\bibinfo  {journal} {New Journal of Physics}\
  }\textbf {\bibinfo {volume} {16}},\ \bibinfo {pages} {113063} (\bibinfo
  {year} {2014})}\BibitemShut {NoStop}%
\bibitem [{\citenamefont {De~Domenico}\ \emph
  {et~al.}(2015{\natexlab{b}})\citenamefont {De~Domenico}, \citenamefont
  {Sol\'{e}-Ribalta}, \citenamefont {Omodei}, \citenamefont {G\'{o}mez},\ and\
  \citenamefont {Arenas}}]{dedom2015}%
  \BibitemOpen
  \bibfield  {author} {\bibinfo {author} {\bibfnamefont {M.}~\bibnamefont
  {De~Domenico}}, \bibinfo {author} {\bibfnamefont {A.}~\bibnamefont
  {Sol\'{e}-Ribalta}}, \bibinfo {author} {\bibfnamefont {E.}~\bibnamefont
  {Omodei}}, \bibinfo {author} {\bibfnamefont {S.}~\bibnamefont {G\'{o}mez}}, \
  and\ \bibinfo {author} {\bibfnamefont {A.}~\bibnamefont {Arenas}},\
  }\href@noop {} {\bibfield  {journal} {\bibinfo  {journal} {Nature
  Communications}\ }\textbf {\bibinfo {volume} {6}},\ \bibinfo {pages} {6868}
  (\bibinfo {year} {2015}{\natexlab{b}})}\BibitemShut {NoStop}%
\bibitem [{\citenamefont {Sol\'{e}-Ribalta}\ \emph {et~al.}(2016)\citenamefont
  {Sol\'{e}-Ribalta}, \citenamefont {De~Domenico}, \citenamefont {G{\'o}mez},\
  and\ \citenamefont {Arenas}}]{sole2016random}%
  \BibitemOpen
  \bibfield  {author} {\bibinfo {author} {\bibfnamefont {A.}~\bibnamefont
  {Sol\'{e}-Ribalta}}, \bibinfo {author} {\bibfnamefont {M.}~\bibnamefont
  {De~Domenico}}, \bibinfo {author} {\bibfnamefont {S.}~\bibnamefont
  {G{\'o}mez}}, \ and\ \bibinfo {author} {\bibfnamefont {A.}~\bibnamefont
  {Arenas}},\ }\href@noop {} {\bibfield  {journal} {\bibinfo  {journal}
  {Physica D}\ }\textbf {\bibinfo {volume} {323}},\ \bibinfo {pages} {73}
  (\bibinfo {year} {2016})}\BibitemShut {NoStop}%
\bibitem [{\citenamefont {Taylor}\ \emph
  {et~al.}(2017{\natexlab{b}})\citenamefont {Taylor}, \citenamefont {Myers},
  \citenamefont {Clauset}, \citenamefont {Porter},\ and\ \citenamefont
  {Mucha}}]{taylor2017eigenvector}%
  \BibitemOpen
  \bibfield  {author} {\bibinfo {author} {\bibfnamefont {D.}~\bibnamefont
  {Taylor}}, \bibinfo {author} {\bibfnamefont {S.~A.}\ \bibnamefont {Myers}},
  \bibinfo {author} {\bibfnamefont {A.}~\bibnamefont {Clauset}}, \bibinfo
  {author} {\bibfnamefont {M.~A.}\ \bibnamefont {Porter}}, \ and\ \bibinfo
  {author} {\bibfnamefont {P.~J.}\ \bibnamefont {Mucha}},\ }\href@noop {}
  {\bibfield  {journal} {\bibinfo  {journal} {Multiscale Modeling \&
  Simulation}\ }\textbf {\bibinfo {volume} {15}},\ \bibinfo {pages} {537}
  (\bibinfo {year} {2017}{\natexlab{b}})}\BibitemShut {NoStop}%
\bibitem [{\citenamefont {DeFord}\ and\ \citenamefont
  {Pauls}(2018)}]{deford2018new}%
  \BibitemOpen
  \bibfield  {author} {\bibinfo {author} {\bibfnamefont {D.~R.}\ \bibnamefont
  {DeFord}}\ and\ \bibinfo {author} {\bibfnamefont {S.~D.}\ \bibnamefont
  {Pauls}},\ }\href@noop {} {\bibfield  {journal} {\bibinfo  {journal} {Journal
  of Complex Networks}\ }\textbf {\bibinfo {volume} {6}},\ \bibinfo {pages}
  {353} (\bibinfo {year} {2018})}\BibitemShut {NoStop}%
\bibitem [{\citenamefont {Taylor}\ \emph
  {et~al.}(2019{\natexlab{a}})\citenamefont {Taylor}, \citenamefont {Porter},\
  and\ \citenamefont {Mucha}}]{taylor2019supracentrality}%
  \BibitemOpen
  \bibfield  {author} {\bibinfo {author} {\bibfnamefont {D.}~\bibnamefont
  {Taylor}}, \bibinfo {author} {\bibfnamefont {M.~A.}\ \bibnamefont {Porter}},
  \ and\ \bibinfo {author} {\bibfnamefont {P.~J.}\ \bibnamefont {Mucha}},\
  }\enquote {\bibinfo {title} {Supracentrality analysis of temporal networks
  with directed interlayer coupling},}\ in\ \href {\doibase
  10.1007/978-3-030-23495-9_17} {\emph {\bibinfo {booktitle} {Temporal Network
  Theory}}},\ \bibinfo {editor} {edited by\ \bibinfo {editor} {\bibfnamefont
  {P.}~\bibnamefont {Holme}}\ and\ \bibinfo {editor} {\bibfnamefont
  {J.}~\bibnamefont {Saram{\"a}ki}}}\ (\bibinfo  {publisher} {Springer
  International Publishing},\ \bibinfo {year} {2019})\ pp.\ \bibinfo {pages}
  {325--344}\BibitemShut {NoStop}%
\bibitem [{\citenamefont {Taylor}\ \emph
  {et~al.}(2019{\natexlab{b}})\citenamefont {Taylor}, \citenamefont {Porter},\
  and\ \citenamefont {Mucha}}]{taylor2019tunable}%
  \BibitemOpen
  \bibfield  {author} {\bibinfo {author} {\bibfnamefont {D.}~\bibnamefont
  {Taylor}}, \bibinfo {author} {\bibfnamefont {M.~A.}\ \bibnamefont {Porter}},
  \ and\ \bibinfo {author} {\bibfnamefont {P.~J.}\ \bibnamefont {Mucha}},\
  }\href@noop {} {\bibfield  {journal} {\bibinfo  {journal} {arXiv preprint
  arXiv:1904.02059}\ } (\bibinfo {year} {2019}{\natexlab{b}})}\BibitemShut
  {NoStop}%
\bibitem [{\citenamefont {De~Domenico}\ \emph {et~al.}(2016)\citenamefont
  {De~Domenico}, \citenamefont {Granell}, \citenamefont {Porter},\ and\
  \citenamefont {Arenas}}]{de2016physics}%
  \BibitemOpen
  \bibfield  {author} {\bibinfo {author} {\bibfnamefont {M.}~\bibnamefont
  {De~Domenico}}, \bibinfo {author} {\bibfnamefont {C.}~\bibnamefont
  {Granell}}, \bibinfo {author} {\bibfnamefont {M.~A.}\ \bibnamefont {Porter}},
  \ and\ \bibinfo {author} {\bibfnamefont {A.}~\bibnamefont {Arenas}},\
  }\href@noop {} {\bibfield  {journal} {\bibinfo  {journal} {Nature Physics}\
  }\textbf {\bibinfo {volume} {12}},\ \bibinfo {pages} {901} (\bibinfo {year}
  {2016})}\BibitemShut {NoStop}%
\bibitem [{\citenamefont {Taylor}()}]{daneGithub}%
  \BibitemOpen
  \bibfield  {author} {\bibinfo {author} {\bibfnamefont {D.}~\bibnamefont
  {Taylor}},\ }\href@noop {} {\enquote {\bibinfo {title}
  {\url{https://github.com/taylordr/multiplexMarkovChains}},}\ }\BibitemShut
  {NoStop}%
\bibitem [{\citenamefont {Skardal}\ \emph {et~al.}(2014)\citenamefont
  {Skardal}, \citenamefont {Taylor},\ and\ \citenamefont
  {Sun}}]{skardal2014optimal}%
  \BibitemOpen
  \bibfield  {author} {\bibinfo {author} {\bibfnamefont {P.~S.}\ \bibnamefont
  {Skardal}}, \bibinfo {author} {\bibfnamefont {D.}~\bibnamefont {Taylor}}, \
  and\ \bibinfo {author} {\bibfnamefont {J.}~\bibnamefont {Sun}},\ }\href@noop
  {} {\bibfield  {journal} {\bibinfo  {journal} {Physical Review Letters}\
  }\textbf {\bibinfo {volume} {113}},\ \bibinfo {pages} {144101} (\bibinfo
  {year} {2014})}\BibitemShut {NoStop}%
\bibitem [{\citenamefont {Taylor}\ \emph
  {et~al.}(2016{\natexlab{b}})\citenamefont {Taylor}, \citenamefont {Skardal},\
  and\ \citenamefont {Sun}}]{taylor2016synchronization}%
  \BibitemOpen
  \bibfield  {author} {\bibinfo {author} {\bibfnamefont {D.}~\bibnamefont
  {Taylor}}, \bibinfo {author} {\bibfnamefont {P.~S.}\ \bibnamefont {Skardal}},
  \ and\ \bibinfo {author} {\bibfnamefont {J.}~\bibnamefont {Sun}},\
  }\href@noop {} {\bibfield  {journal} {\bibinfo  {journal} {SIAM Journal on
  Applied Mathematics}\ }\textbf {\bibinfo {volume} {76}},\ \bibinfo {pages}
  {1984} (\bibinfo {year} {2016}{\natexlab{b}})}\BibitemShut {NoStop}%
\bibitem [{\citenamefont {Bapat}\ and\ \citenamefont
  {Raghavan}(1997)}]{bapat1997}%
  \BibitemOpen
  \bibfield  {author} {\bibinfo {author} {\bibfnamefont {R.~B.}\ \bibnamefont
  {Bapat}}\ and\ \bibinfo {author} {\bibfnamefont {T.~E.~S.}\ \bibnamefont
  {Raghavan}},\ }\href@noop {} {\emph {\bibinfo {title} {Nonnegative Matrices
  and Applications}}},\ Vol.~\bibinfo {volume} {64}\ (\bibinfo  {publisher}
  {Cambridge University Press},\ \bibinfo {address} {Cambridge,UK},\ \bibinfo
  {year} {1997})\BibitemShut {NoStop}%
\bibitem [{\citenamefont {Sahoo}\ and\ \citenamefont
  {Riedel}(1998)}]{sahoo1998mean}%
  \BibitemOpen
  \bibfield  {author} {\bibinfo {author} {\bibfnamefont {P.}~\bibnamefont
  {Sahoo}}\ and\ \bibinfo {author} {\bibfnamefont {T.}~\bibnamefont {Riedel}},\
  }\href@noop {} {\emph {\bibinfo {title} {Mean Value Theorems and Functional
  Equations}}}\ (\bibinfo  {publisher} {World Scientific},\ \bibinfo {year}
  {1998})\BibitemShut {NoStop}%
\bibitem [{\citenamefont {Kato}(2013)}]{kato2013perturbation}%
  \BibitemOpen
  \bibfield  {author} {\bibinfo {author} {\bibfnamefont {T.}~\bibnamefont
  {Kato}},\ }\href@noop {} {\emph {\bibinfo {title} {Perturbation Theory for
  Linear Operators}}},\ Vol.\ \bibinfo {volume} {132}\ (\bibinfo  {publisher}
  {Springer Science \& Business Media},\ \bibinfo {year} {2013})\BibitemShut
  {NoStop}%
\end{thebibliography}%

\end{document}